\documentclass[runningheads]{llncs}
\usepackage[T1]{fontenc}
\usepackage{booktabs}
\usepackage{framed}
\usepackage{amsfonts}
\usepackage{amssymb}
\usepackage{pgfplots}
\usepackage{pgfplotstable}
\usepackage{longtable} 
\usepackage{multicol} 
\usepackage{array}  
\usepackage{ifthen} 
\usepackage{amsmath} 

\usepackage{graphicx}
\begin{document}
%
\title{In-context learning for the classification of manipulation techniques in phishing emails}
%
%
\author{Antony Dalmiere\inst{1}\orcidID{0009-0009-0019-112X} \and 
Guillaume Auriol\inst{1,2}\orcidID{0009-0001-2775-5345} \and
Vincent Nicomette\inst{1,2}\orcidID{0000-0001-9482-004X} \and Pascal Marchand\inst{3}\orcidID{0000-0002-4190-8213}}
\authorrunning{Dalmiere et al.}
%
\institute{CNRS, LAAS, 7 avenue du colonel Roche, F-31400,\\ \email{[firstname.lastname]@laas.fr}\and
 Université de Toulouse, INSA, LAAS, F-31400\\
\and
LERASS, Université of Toulouse, France\\
\email{pascal.marchand@iut-tlse3.fr}}
\maketitle              
\begin{abstract}
Traditional phishing detection often overlooks psychological manipulation. This study investigates using Large Language Model (LLM) In-Context Learning (ICL) for fine-grained classification of phishing emails based on a taxonomy of 40 manipulation techniques. Using few-shot examples with GPT-4o-mini on real-world French phishing emails (SignalSpam), we evaluated performance against a human-annotated test set (100 emails). The approach effectively identifies prevalent techniques (e.g., Baiting, Curiosity Appeal, Request For Minor Favor) with a promising accuracy of 0.76. This work demonstrates ICL's potential for nuanced phishing analysis and provides insights into attacker strategies.

\keywords{Phishing \and Large Language Models \and In-Context Learning \and Psychological Manipulation \and Cybersecurity \and Natural Language Processing.}
\end{abstract}
\section{Introduction}
Phishing attacks have become a sophisticated and widespread threat, posing significant risks to individuals, organizations, and economies globally \cite{herley2009a}. These attacks, often disguised as legitimate communications, exploit human vulnerabilities to steal sensitive information, financial assets, and intellectual property \cite{jagatic2007a,sheng2010a}. Traditional detection methods, primarily focusing on technical indicators such as IP addresses, URLs, and domain names, have proven insufficient in combating the increasingly sophisticated tactics employed by phishers \cite{fette2007a,garera2007a,whittaker2010a}. These technical approaches often fail to capture the nuances of social engineering and psychological manipulation, leaving a critical gap in our ability to effectively identify and prevent phishing attacks \cite{lee2010a,sheng2007a}. The limitation of these methods is further exacerbated by the lack of a deep understanding of the psychological mechanisms that make phishing emails so effective \cite{cialdini2001a,kahneman2011a,li2020a}. This limitation highlights the urgent need for approaches that can analyze the content of phishing emails from a psychological perspective, identifying the subtle manipulation techniques that are used to deceive recipients and trigger compliance \cite{tian2024a,ovelgonne2017a,workman2008a}.

The existing body of research in phishing detection often overlooks the critical role of psychological manipulation \cite{caputo2014a,ferreira2015a}. While the persuasive elements of phishing messages has been studied \cite{kim2013a}, they have not fully incorporated a granular analysis of specific psychological manipulation techniques or leveraged the power of Large Language Models (LLMs) for advanced classification. This study addresses this gap by investigating an approach that leverages the in-context learning (ICL) capabilities of LLMs to classify phishing emails based on a detailed taxonomy of distinct psychological manipulation techniques.
These techniques include, but are not limited to, appeals to authority, scarcity, curiosity, and emotional triggers such as guilt and fear \cite{cialdini2009a,goldstein2008a,pratkanis2007a}. By focusing on the psychological underpinnings of phishing attacks, we aim to provide a more nuanced and effective approach to detection and prevention.

In-context learning offers a promising avenue for this task because it enables LLMs to classify text based on examples provided in the prompt, without requiring explicit fine-tuning on labeled data \cite{winata2021,dong2020}. This ability is particularly beneficial for phishing detection, where new manipulation techniques emerge regularly, and explicit training on each new tactic is impractical. By providing carefully crafted examples of each manipulation technique within the prompt, the LLM can learn to identify these techniques in unseen phishing emails, even when they are presented in novel contexts and combinations. This approach allows for a more flexible and adaptable detection system that can respond to the ever-changing tactics of phishers.
This paper describes an experiment carried out on real data, consisting of phishing emails collected in 2023 by SignalSpam\footnote[1]{https://www.signal-spam.fr/}, a French organization dedicated to the collection and analysis of French phishing emails.
We leverage ICL with GPT-4o-mini LLM to classify these phishing emails based on various specific manipulation techniques.

To summarize, the key contributions of this paper are:
1) an empirical study of the effectiveness of in-context learning for accurately identifying manipulation techniques in phishing emails; and 2) a detailed analysis of occurences of manipulation techniques that are prevalent in real-world phishing emails.


The paper is organized as follows. Section \ref{sec:related} describes related work. Section \ref{sec:overview} presents an overview of the approach proposed in this paper. Section \ref{sec:methodo} describes the methodology followed: the data considered, the in-context learning setup and experiments setup. Section \ref{sec:results} presents the results of the experiments. Section \ref{sec:discussion} proposes a discussion and Section \ref{sec:conclusion} concludes the paper.

\section{Related Works}
\label{sec:related}

Early approaches to phishing detection laid the groundwork using rule-based systems \cite{basnet2011a} and initial machine learning classifiers \cite{fette2007aa}. Basnet et al. \cite{basnet2011a} effectively employed rule-based methods to identify phishing webpages, focusing on understandable rules derived from observable patterns. Fette et al. \cite{fette2007aa} pioneered machine learning for phishing email detection, achieving high accuracy using features extracted from email content. While these works were crucial in establishing automated detection, they inherently focused on binary classification (phishing vs. not phishing) and primarily leveraged technical and surface-level content features, lacking deeper insights into the psychological aspects of phishing.

Building upon these foundations, subsequent research utilized more sophisticated machine learning and NLP techniques, still largely within the binary classification paradigm. Gikandi et al. \cite{gikandi2024aa} explored sentence-level analysis with KNN for improved detection accuracy, and Noah et al. \cite{noah2022aa} developed PhisherCop, achieving high accuracy using SGD and SVC classifiers. These studies demonstrated advancements in feature engineering and classification algorithms, yet their primary objective remained binary phishing email detection, without explicitly addressing the spectrum of psychological manipulation tactics. Systematic reviews by Kyaw et al. \cite{kyaw2024aa} and Salloum et al. \cite{salloum2022a} further highlight the field's progression towards deep learning and NLP for phishing detection, confirming the trend towards sophisticated technical analysis, but also implicitly revealing a persistent gap in granular, psychologically-informed classification.

The emergence of Large Language Models (LLMs) has introduced new capabilities to the field. Koide et al. \cite{koide2024aa} in ChatSpamDetector showcased GPT-4's remarkable accuracy in identifying phishing emails using In Context Learning, emphasizing the potential of LLMs for advanced contextual interpretation, cutting the cost of classification to only inferance. 

While these studies signal a shift towards leveraging LLMs' sophisticated text understanding, they, like earlier works, primarily evaluate performance in terms of binary classification—identifying if an email is phishing, rather than how it manipulates the recipient at a cognitive level. Pan et al. \cite{pan2024aa} explored LLMs' zero-shot learning for fallacy classification, demonstrating their capacity for classification on social media. Shahriar et al. \cite{shahriar2022aa} made a significant step by incorporating psychological trait scoring into phishing detection, using traits like urgency and fear to enhance classification. However, even this approach, while psychologically informed, operates at a higher level of abstraction, classifying emails based on the presence of too broad psychological traits for binary classification, rather than dissecting the specific techniques of manipulation.

Our work addresses this gap by moving beyond binary
phishing/safe determinations and broad psychological traits. We introduce a novel methodology that harnesses the in-context learning capabilities of LLMs to achieve a fine-grained classification of
phishing emails, according to a taxonomy of distinct
psychological manipulation techniques, enabling a level of nuanced analysis previously unexplored. This allows for a deeper understanding of the specific cognitive exploits used in phishing attacks, offering a significant advancement over existing binary classification approaches and even trait-based psychological scoring.

\section{Approach overview}
\label{sec:overview}

\begin{figure}[h!]
    \centering
    \includegraphics[width=0.75\textwidth]{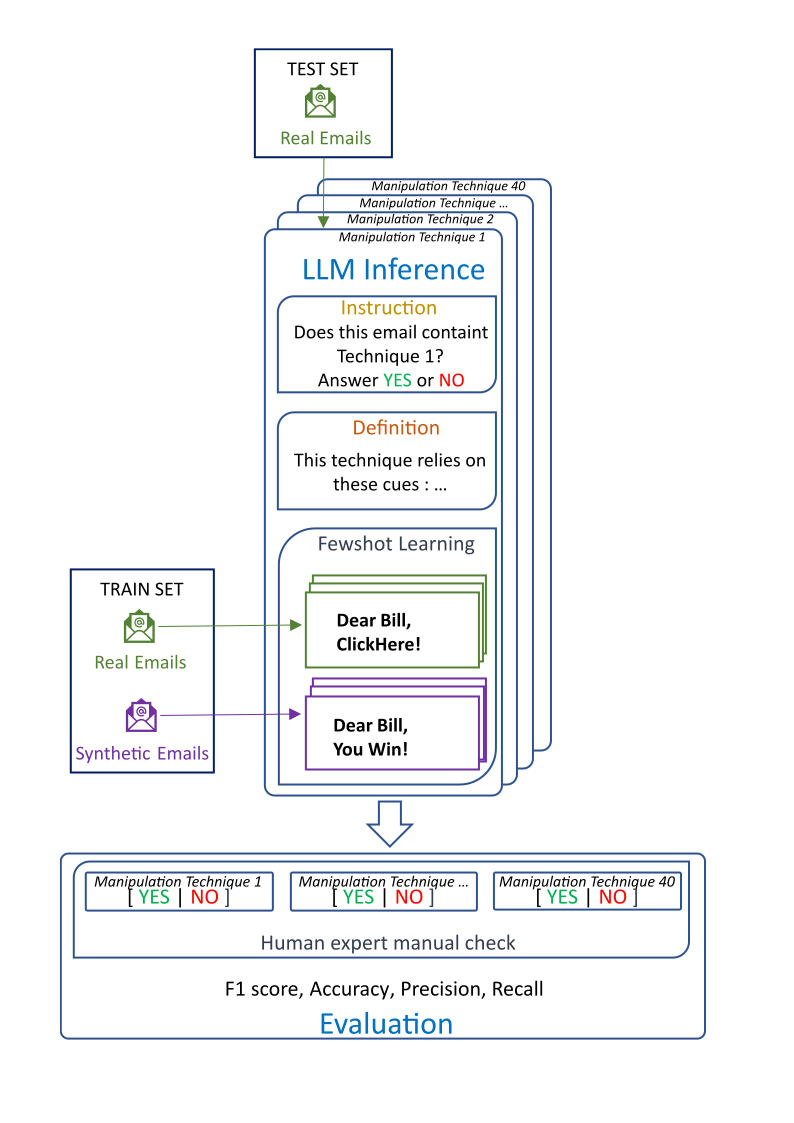}
    \caption{Overview of the In-Context Learning (ICL) Methodology for Psychological Manipulation Technique Classification. 
    Instruction, definition and exemples are given for illustration and are detailled after.} 
    \label{fig:methodology_overview}
\end{figure}

Figure \ref{fig:methodology_overview} illustrates our In-Context Learning (ICL) methodology for classifying psychological manipulation techniques in phishing emails using language models (LLMs), specifically GPT-4o-mini as detailed in Section \ref{sec:methodo}. The process involves several stages, starting with dataset preparation (Section \ref{sec:methodo}). This includes a train set containing real emails manually labelized by an expert (providing numerous \textbf{samples}, i.e., instances of techniques) and synthetic email examples (see Annex \ref{annex:synthetic}) to ensure coverage of all 40 techniques (listed in Annex \ref{annex:techniques}). A separate 100 emails test set is used for evaluation. These datasets serve as input for the LLM inference stage. Here, for each technique, the LLM analyzes test emails using a structured prompt (see Annex \ref{annex:systempromptclassification}). This prompt includes: an {\em Instruction} stating the binary classification task ("Does this email contain this technique?"), a precise academic {\em Definition} of the technique, and {\em Few-shot Learning} examples (e.g., "Hello, click here!", "Win 10\$") from the train set to guide the classification. Based on this prompt, the LLM determines the presence or absence of the technique in the test email. Following the model's initial classification, a manual check stage is performed. A human expert reviews the LLM's labels on the 100 emails test set to ensure reliability, particularly for subtle techniques like \textbf{Reciprocity} or \textbf{Baiting}. Finally, the evaluation stage rigorously assesses the methodology's performance using standard metrics (F1, Accuracy, Precision, Recall, detailed in Section \ref{sec:methodo}) and Average Weighted Accuracy, evaluating the model's ability to accurately identify manipulation techniques within the emails.

\section{Methodology}
\label{sec:methodo}

The methodology we used encompasses the definition of specific classification categories, the selection of a relevant dataset, the design of effective ICL prompts and the implementation of a comprehensive evaluation process.

\subsection{Classification of manipulation techniques}

This study aims at automatically classifying phishing emails according to a taxonomy of 40 distinct psychological manipulation techniques described in \cite{dalmiere:hal-05027416}. The detailed description of these techniques is out of the scope of this paper but a brief description of the most relevant for this paper is provided 
in Appendix \ref{annex:techniques}. Just to give an idea to the reader of the variety of these techniques, a few examples are provided below: 
\begin{itemize}
    \item \textbf{Authority}: Enhances credibility by associating the message with a seemingly reliable or legitimate source
    \item \textbf{Baiting}: Uses positive emotional stimuli, such as offers of rewards, money, or romantic promises, to lure targets and direct their attention.
    \item \textbf{Contact Data Present}: Incorporates contact details (phone numbers, email addresses) to falsely signal legitimacy and market orientation. 
    \item \textbf{Curiosity Appeal}: Increases  motivation to engage by piquing the target's innate curiosity. 
    \item \textbf{Flattery}:  Influences by offering compliments and praise.
    \item \textbf{Request For Minor Favor}: Seeks small, seemingly harmless acts of compliance initially (micro-commitments). 
    \item \textbf{Time Pressure}: Reduces cognitive processing and critical evaluation time by creating a sense of urgency and time sensitivity. 
\end{itemize}

\ifthenelse{\boolean{false}}
{
\begin{itemize}
    \item \textbf{Argumentative Mille Feuille}: Overwhelms the target with a high volume of arguments, regardless of individual argument strength. 
    \item \textbf{Attractiveness}: Utilizes physical appeal to build trust rapidly.
    \item \textbf{Authority}: Enhances credibility by associating the message with a seemingly reliable or legitimate source
    \item \textbf{Baiting}: Uses positive emotional stimuli, such as offers of rewards, money, or romantic promises, to lure targets and direct their attention.
    \item \textbf{Contact Data Present}: Incorporates contact details (phone numbers, email addresses) to falsely signal legitimacy and market orientation. 
    \item \textbf{Curiosity Appeal}: Increases  motivation to engage by piquing the target's innate curiosity. 
    \item \textbf{Disgust Calling}: Uses repulsive stimuli (images, descriptions) to induce the emotion of disgust.
    \item \textbf{Door In The Face}: Secures compliance to a smaller, target request by first presenting an unreasonably large, initial request that is likely to be refused.
    \item \textbf{Either Or Fallacy}: Restricts perceived options to a false dichotomy, presenting only two opposing choices.
    \item \textbf{Fake Consistency}: Tricks the target into behaving consistently with a fabricated past or present commitment or belief.
    \item \textbf{Fake Divergence}: Creates an illusion of widespread disagreement or uncertainty on a topic to undermine the target's confidence in their existing knowledge. 
    \item \textbf{False Impartiality}: Presents the manipulator as neutral and disinterested to enhance credibility. 
    \item \textbf{Flattery}:  Influences by offering compliments and praise.
    \item \textbf{Foot In The Door}: Achieves compliance with a larger, ultimate request by first securing agreement to a smaller, initial request.
    \item \textbf{Group Thinking}: Leverages the desire for group consensus and conformity to suppress dissenting opinions within a group setting. 
    \item \textbf{Guilt Calling}: Manipulates through remorse, exploiting internalized guilt arising from perceived transgressions.
    \item \textbf{Humanitarianism}: Exploits guilt by portraying oneself as a victim in need of help to evoke sympathy and compassion.
    \item \textbf{Hyperchleuasm}: Uses exaggeration or irony to inoculate against reasonable counterarguments by presenting an extreme version.
    \item \textbf{Imply}: Subtly influences through indirect suggestion rather than explicit commands. 
    \item \textbf{Humour}: Distracts attention and reduces message comprehension by employing humor. 
    \item \textbf{Impress Management}: Enhances persuasiveness by projecting positive traits, making the source more appealing and trustworthy during heuristic processing.
    \item \textbf{Indignation}: Evokes moderate anger at perceived injustice or violation of values, increasing motivation to act against the perceived wrongdoer. 
    \item \textbf{Include Half Confidential Details}: Builds a false sense of trust and rapport by including seemingly confidential or personalized details.
    \item \textbf{Metaphor}: Uses figurative language (metaphors) to subtly direct interpretation and frame understanding in a specific, manipulator-preferred way. 
    \item \textbf{Non Verbal Synchronocity}: Enhances message reception and persuasiveness through synchronized non-verbal cues, improving comprehension and agreement during real-time interaction.
    \item \textbf{Obscurantism}: Employs complex jargon and obfuscated language to deceive or impress. 
    \item \textbf{Parallel Tasking}: Reduces attentional capacity and cognitive resources by introducing a secondary task to be performed concurrently with processing the manipulative message.
    \item \textbf{Provocation}: Induces strong anger to heighten motivation to engage, but simultaneously diminishing critical thinking and increasing impulsive behavior directed at the perceived source of provocation. 
    \item \textbf{Reciprocity}:  Obliges the target to reciprocate after receiving a favor or perceived benefit.
    \item \textbf{Request For Minor Favor}: Seeks small, seemingly harmless acts of compliance initially (micro-commitments). 
    \item \textbf{Reverse Psychology}: Enhances psychological reactance by forbidding an action.
    \item \textbf{Scarcity}: Inflates perceived value and urgency by emphasizing rarity and limited availability. 
    \item \textbf{Secure Communication Channel}: Leverages trusted communication mediums to falsely assure targets of security and legitimacy. 
    \item \textbf{Semantic Attack}: Employs visually similar domain names, URLs, or email addresses to mimic legitimate entities and deceive users.
    \item \textbf{Shameful Disclosure}: Manipulates using the threat of revealing embarrassing or norm-violating behavior. 
    \item \textbf{Similarity}: Establishes perceived shared traits or characteristics to build trust and rapport rapidly.
    \item \textbf{Social Pressure}: Influences by demonstrating that others (can be a fabricated majority) are already complying or adopting a certain behavior. 
    \item \textbf{Threatening}: Instills fear by conveying potential negative consequences. 
    \item \textbf{Time Pressure}: Reduces cognitive processing and critical evaluation time by creating a sense of urgency and time sensitivity. 
    \item \textbf{Use Of Statistics}: Employs seemingly objective numerical data and statistics to sway opinions and create a sense of certainty and validity. 
\end{itemize}
}

Each technique has a clear definition, 
grounded in psychological literature and contextualized for phishing attacks. These cognitively-focused definitions were utilized by the human expert annotator during the dataset labeling verification process, ensuring a robust and psychologically grounded classification.

\subsection{Dataset}
\label{subsec:dataset}
\ifthenelse{\boolean{false}}
{
\subsection{Dataset}
\begin{figure}[h!]
    \centering
    \includegraphics[width=0.8\textwidth]{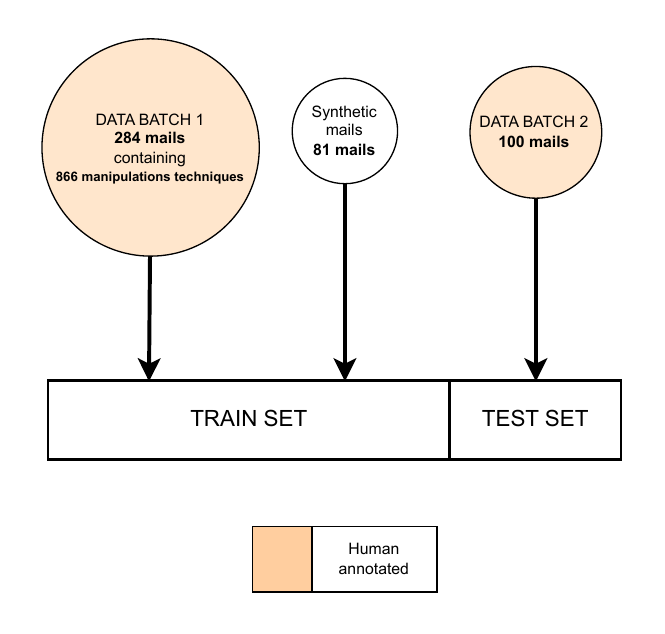}
    \caption{Dataset composition.} 
    \label{fig:datacomposition}
\end{figure}
}
The dataset for this study was provided by SignalSpam, a French organization dedicated to collecting and analyzing phishing emails targeting French users. SignalSpam is a source for real-world phishing data in France. 
The train set includes 284 phishing email collected in 2023. These data were manually labelized by an expert to associate each mail with the presence of each technique giving us 866 manipulation techniques samples (each email often carries multiple manipulation techniques). 
After labelizing the train set, if a label appears in the manipulation techniques samples less than 5 times, we completed the train set with 81 synthetic emails (each mail carrying one single manipulation technique), resulting in a whole training set of 947 manipulation techniques samples. Appendix \ref{annex:synthetic} details the techniques with their number of real-life examples and synthetic mails.

\ifthenelse{\boolean{false}}
{
\begin{table}[!ht]
\centering
\caption{List of techniques with their number of real-life examples and synthetic mails}
\label{tab:techniques} 
\begin{tabular}{lcc}
\toprule
\textbf{Technique name} & \textbf{Real examples} & \textbf{Synthetic mails} \\
\midrule
Argumentative Mille Feuille & 6 & 0 \\
Attractiveness & 4 & 1 \\
Authority & 9 & 0 \\
Baiting & 160 & 0 \\
Contact Data Present & 142 & 0 \\
Curiosity Appeal & 38 & 0 \\
Disgust Calling & 0 & 5 \\
Door in the Face & 0 & 5 \\
Either or Fallacy & 0 & 5 \\
Fake Consistency & 23 & 0 \\
Fake Divergence & 0 & 5 \\
False Impartiality & 3 & 2 \\
Flattery & 33 & 0 \\
Foot in the Door & 6 & 0 \\
Group Thinking & 0 & 5 \\
Guilt Calling & 26 & 0 \\
Humanitarianism & 0 & 5 \\
Humour & 41 & 0 \\
Hyperchleuasm & 0 & 5 \\
Imply & 0 & 5 \\
Impress Management & 19 & 0 \\
Include Half Confidential Details & 50 & 0 \\
Indignation & 1 & 4 \\
Metaphor & 13 & 0 \\
Non-Verbal Synchronicity & 0 & 5 \\
Obscurantism & 17 & 0 \\
Parallel Tasking & 3 & 2 \\
Provocation & 1 & 4 \\
Reciprocity & 0 & 5 \\
Request for Minor Favor & 52 & 0 \\
Reverse Psychology & 0 & 5 \\
Scarcity & 16 & 0 \\
Secure Communication Channel & 7 & 0 \\
Semantic Attack & 95 & 0 \\
Shameful Disclosure & 0 & 5 \\
Similarity & 0 & 5 \\
Social Pressure & 2 & 3 \\
Threatening & 54 & 0 \\
Time Pressure & 38 & 0 \\
Use of Statistics & 7 & 0 \\
\midrule
\textbf{Total} & \textbf{866} & \textbf{81} \\
\bottomrule
\end{tabular}
\end{table}
}


The test set consists in a random sample of 4000 classification YES/NO of 40 manipulation techniques by labeling these techniques over 100 mails (different from the mails used for the train set). This test set is automatically labeled by the GPT-4o-mini Large Language Model using in-context learning (as described in next Subsection), for each email and for each manipulation technique.

Establishing ground truth for 40 psychological manipulation techniques in phishing demanded rigorous interdisciplinary integration between cognitive science and cybersecurity. Initial assessments highlighted the subtle nuances differentiating techniques (e.g., {\bf Foot in the Door} vs. {\bf Request for Minor Favor}) and the inherent variability in multi-annotator agreement for such a specialized taxonomy. To achieve the highest interpretive consistency and accuracy, we strategically adopted a single-expert validation protocol for the test set, ensuring psychological grounding and fair evaluation. An LLM provided baseline annotations, meticulously reviewed by an expert uniquely qualified in both cognitive science and cybersecurity. This focused approach yields a coherent, specialized, and psychologically-grounded benchmark essential for evaluating model performance on this complex task.
We are conscious that a bigger test set would be better to improve the analysis but this number was chosen because  labeling over 40 techniques is very time-consuming for the expert (consists of 4000 evaluations for the expert). This human-annotated subset was used to compute the evaluation metrics presented further in the results section. Preprocessing was performed on the email content. The preprocessing consists of extracting the text from the raw email. Email attachment filenames and subjects were not passed along with the email’s body to avoid missing semantics carried only in filenames or subjects. All other information in the email, such as headers, was discarded to guarantee anonymization of the email; moreover, it does not carry semantics.

\subsection{In-Context Learning Setup}

This study employs the in-context learning capabilities of state-of-the-art Large Language Models. Specifically, all experiments were conducted using the GPT-4o-mini model. This model was selected based on its superior performance, achieving the highest Average Weighted Accuracy in our benchmark evaluations. 
Section~\ref{sec:comparison} presents a comparative analysis of nine state-of-the-art LLMs.


The ICL setup involves carefully crafting prompts that include examples of pre-labeled distinct phishing emails of the training set, labeled with their corresponding manipulation techniques. The prompt is in Appendix \ref{annex:systempromptclassification} and follows the structure:
\begin{enumerate}
    \item \textbf{Instruction}: A brief description of the task, requesting the LLM to identify the manipulation technique used in the email.
    \item \textbf{Definition}: An academic definition of each manipulation technique 
    derived from Dalmiere et al. \cite{dalmiere:hal-05027416}.
    \item \textbf{Examples}: A set of few-shot examples, which include a phishing email and its corresponding manipulation technique. The number of examples was varied in different experiments to evaluate the impact of the number of examples on performance. Initially, these examples were 
    taken from the training set. To ensure a sufficient number of examples for each manipulation technique, we aimed for a minimum of five examples per technique. In cases where the pre-labeling yielded fewer than five real-world examples for a specific technique, we employed a synthetic data augmentation approach. Using GPT-4o, we generated synthetic examples of phishing emails that exemplified the target manipulation technique \cite{long2024,evanko2021gpt3,Whitfield_2021}. Prompt used is detailed in Annex \ref{annex:prompt}. The number of synthetic emails for each manipulation technique is provided in Annex \ref{annex:synthetic}. This ensured that each technique was represented by at least five examples in the ICL prompts. For instance, techniques like {\bf Disgust Calling} and {\bf Door In The Face} initially had 0 real-world examples, while {\bf Attractiveness} had only 4. In such cases, synthetic examples were crucial to meet the minimum example requirement.
    \item \textbf{Query}: The phishing email for which the LLM needs to identify the manipulation technique.
\end{enumerate}

The prompt design was carefully considered to provide the LLMs with clear instructions and diverse examples of each manipulation technique. The prompts were designed to be clear and concise, minimizing ambiguity and maximizing the likelihood of accurate classification.

\subsection{Evaluation Metrics}
The performance of the ICL models was evaluated using a range of standard metrics:
\begin{itemize}
    \item True Positives (TP): The number of phishing emails correctly classified as containing a specific manipulation technique.
    \item True Negatives (TN): The number of phishing emails correctly classified as not containing a specific manipulation technique.
    \item False Positives (FP): The number of phishing emails incorrectly classified as containing a specific manipulation technique.
    \item False Negatives (FN): The number of phishing emails incorrectly classified as not containing a specific manipulation technique.
    \item Accuracy (Acc) : The overall correctness of the classification: \\(TP+TN)/(TP+FN+FP+FN)
    \item Recall (Rec): The ability of the model to identify all relevant instances of a specific manipulation technique: TP/(TP+FN)
    \item Precision (Prec): The accuracy of the model in classifying an email as containing a specific manipulation technique: TP/(TP+FP)
    \item F1-score (F1) : The harmonic mean of precision and recall, providing a balanced measure of the model's performance : 2*(Prec * Rec)/(Prec + Rec)
\end{itemize}

These metrics provide a comprehensive view of the model's performance in classifying the presence of each manipulation techniques in emails. On top of that we added one metric, the Average Weighted Accuracy, which aggregates the accuracy of the classifier over enough represented techniques
calculated by taking the average of the accuracy for each technique, weighted by the number of instances of each technique.
     \[
    \text{Average Weighted Accuracy} = \frac{\sum_{i=1}^{k} (n_i \times \text{Accuracy}_i)}{\sum_{i=1}^{k} n_i}
    \]
    where \( n_i \) is the number of instances for technique \( i \) in testset, \( \text{Accuracy}_i \) is the accuracy for technique \( i \) and \( k \) the number of technique with enough representation in test set.

\subsection{Experimental Setup}


The experimental setup involved running a separate experiment for each of the 40 manipulation techniques. Each experiment consisted of presenting the ICL prompt to the LLMs and evaluating their performance based on the metrics described above. The experiments were conducted once for each technique, with no repeated trials.

\section{Preliminary Results}
\label{sec:results}



\ifthenelse{\boolean{false}}
{
\begin{longtable}{|l|c|c|c|c|c|c|c|c|}
\caption{Detailed quantitative results of our experiments}
\label{tab:Scoretechniques}
\hline
\textbf{Technique} & \textbf{TP} & \textbf{TN} & \textbf{FP} & \textbf{FN} & \textbf{Acc} & \textbf{Rec} & \textbf{Prec} & \textbf{F1} \\
\hline
\endfirsthead
\hline
\textbf{Technique} & \textbf{TP} & \textbf{TN} & \textbf{FP} & \textbf{FN} & \textbf{Acc} & \textbf{Rec} & \textbf{Prec} & \textbf{F1} \\
\hline
\endhead
\hline
\endfoot
\hline
\endlastfoot
\hline
\textit{Highly representative techniques}& & & & & & & & \\
Authority & 6 & 69 & 24 & 1 & 0,75 & 0,86 & 0,20 & 0,32 \\
Argumentative Mille Feuille & 8 & 40 & 49 & 3 & 0,48 & 0,73 & 0,14 & 0,24 \\
Contact Data Present & 33 & 25 & 7 & 35 & 0,58 & 0,49 & 0,82 & 0,61 \\
Curiosity Appeal & 62 & 12 & 17 & 9 & 0,74 & 0,87 & 0,78 & 0,83 \\
Fake Consistency & 42 & 32 & 6 & 20 & 0,74 & 0,68 & 0,88 & 0,76 \\
Foot In The Door & 4 & 74 & 21 & 1 & 0,78 & 0,80 & 0,16 & 0,27 \\
Flattery & 40 & 3 & 57 & 0 & 0,43 & 1,00 & 0,41 & 0,58 \\
Guilt Calling & 3 & 89 & 1 & 7 & 0,92 & 0,30 & 0,75 & 0,43 \\
Humour & 31 & 8 & 61 & 0 & 0,39 & 1,00 & 0,34 & 0,50 \\
Impress Management & 2 & 48 & 47 & 3 & 0,50 & 0,40 & 0,04 & 0,07 \\
Include Half Confidential Details & 14 & 68 & 9 & 9 & 0,82 & 0,61 & 0,61 & 0,61 \\
Metaphor & 26 & 32 & 42 & 0 & 0,58 & 1,00 & 0,38 & 0,55 \\
Obscurantism & 3 & 75 & 15 & 7 & 0,78 & 0,30 & 0,17 & 0,21 \\
Reciprocity & 5 & 86 & 2 & 7 & 0,91 & 0,42 & 0,71 & 0,53 \\
Scarcity & 13 & 66 & 19 & 2 & 0,79 & 0,87 & 0,41 & 0,55 \\
Secure Communication Channel & 5 & 80 & 12 & 3 & 0,85 & 0,62 & 0,29 & 0,40 \\
Request For Minor Favor & 77 & 0 & 22 & 1 & 0,77 & 0,99 & 0,78 & 0,87 \\
Threatening & 14 & 69 & 2 & 15 & 0,83 & 0,48 & 0,88 & 0,62 \\
Time Pressure & 15 & 67 & 16 & 2 & 0,82 & 0,88 & 0,48 & 0,62 \\
Baiting & 57 & 20 & 8 & 15 & 0,77 & 0,79 & 0,88 & 0,83 \\
Semantic Attack & 13 & 64 & 10 & 13 & 0,77 & 0,50 & 0,57 & 0,53 \\
\hline
\textit{Techniques with (TP+FN) less than 5} & & & & & & & & \\
Disgust Calling & 1 & 98 & 1 & 0 & 0,99 & 1,00 & 0,50 & 0,67 \\
Door In The Face & 0 & 99 & 0 & 1 & 0,99 & 0,00 & 0,00 & 0,00 \\
Either Or Fallacy & 0 & 99 & 0 & 1 & 0,99 & 0,00 & 0,00 & 0,00 \\
False Impartiality & 1 & 97 & 2 & 0 & 0,98 & 1,00 & 0,33 & 0,50 \\
Hyperchleuasm & 0 & 99 & 0 & 1 & 0,99 & 0,00 & 0,00 & 0,00 \\
Imply & 0 & 99 & 0 & 1 & 0,99 & 0,00 & 0,00 & 0,00 \\
Parallel Tasking & 2 & 48 & 50 & 0 & 0,50 & 1,00 & 0,04 & 0,07 \\
Provocation & 2 & 60 & 38 & 0 & 0,62 & 1,00 & 0,05 & 0,10 \\
Reverse Psychology & 0 & 99 & 0 & 1 & 0,99 & 0,00 & 0,00 & 0,00 \\
Similarity & 4 & 91 & 5 & 0 & 0,95 & 1,00 & 0,44 & 0,62 \\
Use Of Statitics & 4 & 85 & 11 & 0 & 0,89 & 1,00 & 0,27 & 0,42 \\
\hline
\textit{Techniques with (TP,FN) = (0,0)} & & & & & & & & \\
Attractiveness & 0 & 93 & 7 & 0 & 0,93 & 0,00 & 0,00 & 0,00 \\
Group Thinking & 0 & 99 & 1 & 0 & 0,99 & 0,00 & 0,00 & 0,00 \\
Humanitarianism & 0 & 98 & 2 & 0 & 0,98 & 0,00 & 0,00 & 0,00 \\
Indignation & 0 & 96 & 4 & 0 & 0,96 & 0,00 & 0,00 & 0,00 \\
Non Verbal Synchronicity & 0 & 100 & 0 & 0 & 1,00 & 0,00 & 0,00 & 0,00 \\
Shameful Disclosure & 0 & 99 & 1 & 0 & 0,99 & 0,00 & 0,00 & 0,00 \\
Social Pressure & 0 & 100 & 0 & 0 & 1,00 & 0,00 & 0,00 & 0,00 \\
Fake Divergence & 0 & 99 & 1 & 0 & 0,99 & 0,00 & 0,00 & 0,00 \\
\end{longtable}
}

Table \ref{tab:Scoretechniques} presents the detailed quantitative results of our experiments, including the number of TP, TN, FP, FN, Acc, Rec, Prec, and F1 for each of the 40 manipulation techniques. This table presents the results for the GPT-4O-mini LLM, which provided the best metrics (see Section \ref{sec:comparison}).
This Table 
shows a wide range of performance across the different manipulation techniques. Some techniques, such as {\bf Curiosity Appeal}, {\bf Request For Minor Favor}, and {\bf Baiting}, exhibit relatively high F1-scores, indicating good performance in both recall and precision. Conversely, many techniques, such as {\bf Attractiveness}, {\bf Door In The Face}, and {\bf Group Thinking}, have F1-scores of 0, indicating that the models were unable to classify these techniques effectively.

\footnotesize{
\begin{longtable}[!t]{|l|c|c|c|c|c|c|c|c|}
\caption{Quantitative results (* = refusal from the model)}
\label{tab:Scoretechniques} \\
\hline
\textbf{Technique} & \textbf{TP} & \textbf{TN} & \textbf{FP} & \textbf{FN} & \textbf{Acc} & \textbf{Rec} & \textbf{Prec} & \textbf{F1} \\
\hline
\endfirsthead
\hline
\textbf{Technique} & \textbf{TP} & \textbf{TN} & \textbf{FP} & \textbf{FN} & \textbf{Acc} & \textbf{Rec} & \textbf{Prec} & \textbf{F1} \\
\hline
\endhead
\hline
\endfoot
\hline
\endlastfoot
\hline
\textit{Highly representative techniques} & & & & & & & & \\
Argumentation Mille Feuille & 4 & 53 & 36 & 7  & 0.57 & 0.36 & 0.10 & 0.16 \\
Authority & 5 & 71 & 22 & 2  & 0.76 & 0.71 & 0.19 & 0.29 \\
Baiting* & 53 & 26 & 2 & 18 & 0.79 & 0.74 & 0.96 & 0.83 \\
Contact Data Present & 48 & 21 & 11 & 20 & 0.69 & 0.71 & 0.81 & 0.76 \\
Curiosity Appeal & 64 & 11 & 18 & 7  & 0.75 & 0.90 & 0.78 & 0.84 \\
Fake Consistency & 46 & 28 & 10 & 16 & 0.74 & 0.74 & 0.82 & 0.78 \\
Foot In The Door & 4 & 78 & 17 & 1  & 0.82 & 0.80 & 0.19 & 0.31 \\
Flattery & 37 & 54 & 6 & 3  & 0.91 & 0.93 & 0.86 & 0.89 \\
Guilt Calling & 1 & 90 & 0 & 9  & 0.91 & 0.10 & 1.00 & 0.18 \\
Humour & 28 & 52 & 17 & 3  & 0.80 & 0.90 & 0.62 & 0.74 \\
Impress Management & 2 & 54 & 41 & 3  & 0.56 & 0.40 & 0.05 & 0.08 \\
Include Half Confidential Details & 14 & 68 & 9 & 9  & 0.82 & 0.61 & 0.61 & 0.61 \\
Metaphor & 26 & 32 & 42 & 0  & 0.58 & 1.00 & 0.38 & 0.55 \\
Obscurantism & 3 & 70 & 20 & 7 & 0.73 & 0.30 & 0.13 & 0.18 \\
Reciprocity & 5 & 87 & 1 & 7 & 0.92 & 0.42 & 0.83 & 0.56 \\
Request For Minor Favor* & 60 & 13 & 9 & 17 & 0.73 & 0.77 & 0.87 & 0.82 \\
Scarcity & 13 & 56 & 29 & 2  & 0.69 & 0.87 & 0.31 & 0.46 \\
Secure Communication Channel & 5 & 80 & 12 & 3 & 0.85 & 0.62 & 0.29 & 0.40 \\
Semantic Attack & 14 & 59 & 15 & 12 & 0.73 & 0.54 & 0.48 & 0.51 \\
Threatening & 12 & 70 & 1 & 17 & 0.82 & 0.41 & 0.92 & 0.57 \\
Time Pressure & 13 & 67 & 16 & 4  & 0.80 & 0.76 & 0.45 & 0.57 \\
\hline
\textit{Techniques with (TP+FN) less than 5} & & & & & & & & \\
Disgust Calling & 0 & 99 & 0 & 1 & 0.99 & 0.00 & 0.00 & 0.00 \\
Door In The Face & 0 & 99 & 0 & 1 & 0.99 & 0.00 & 0.00 & 0.00 \\
Either Or Fallacy & 0 & 99 & 0 & 1  & 0.99 & 0.00 & 0.00 & 0.00 \\
False Impartiality & 0 & 90 & 9 & 1 & 0.90 & 0.00 & 0.00 & 0.00 \\
Hyperchleuasm & 0 & 99 & 0 & 1  & 0.99 & 0.00 & 0.00 & 0.00 \\
Imply & 0 & 99 & 0 & 1  & 0.99 & 0.00 & 0.00 & 0.00 \\
Parallel Tasking & 2 & 31 & 67 & 0  & 0.33 & 1.00 & 0.03 & 0.06 \\
Provocation & 1 & 64 & 34 & 1 & 0.65 & 0.50 & 0.03 & 0.05 \\
Reverse Psychology & 0 & 99 & 0 & 1  & 0.99 & 0.00 & 0.00 & 0.00 \\
Similarity & 2 & 93 & 3 & 2  & 0.95 & 0.50 & 0.40 & 0.44 \\
Use Of Statistics & 2 & 93 & 3 & 2  & 0.95 & 0.50 & 0.40 & 0.44 \\
\hline
\textit{Techniques with (TP,FN) = (0,0)} & & & & & & & & \\
Attractiveness & 0 & 92 & 8 & 0 & 0.92 & 0.00 & 0.00 & 0.00 \\
Fake Divergence & 0 & 99 & 1 & 0 & 0.99 & 0.00 & 0.00 & 0.00 \\
Group Thinking & 0 & 100 & 0 & 0 & 1.00 & 0.00 & 0.00 & 0.00 \\
Humanitarianism & 0 & 94 & 6 & 0 & 0.94 & 0.00 & 0.00 & 0.00 \\
Indignation & 0 & 93 & 7 & 0  & 0.93 & 0.00 & 0.00 & 0.00 \\
Non Verbal Synchronicity & 0 & 100 & 0 & 0  & 1.00 & 0.00 & 0.00 & 0.00 \\
Shameful Disclosure & 0 & 97 & 3 & 0  & 0.97 & 0.00 & 0.00 & 0.00 \\
Social Pressure & 0 & 100 & 0 & 0  & 1.00 & 0.00 & 0.00 & 0.00 \\
\end{longtable}
}

\normalsize




Analyzing the combined values of True Positives (TP) and False Negatives (FN) in Table \ref{tab:Scoretechniques} provides valuable insights into the real-world usage of different psychological manipulation techniques by social engineers in phishing emails. While the TP value represents the number of emails correctly identified as containing a specific technique, the sum of TP and FN represents the total number of emails in our dataset that actually contained that technique, regardless of whether the model correctly classified them. This analysis moves beyond simply assessing model performance to gauging the actual practical application of these techniques in social engineering campaigns.
\ifthenelse{\boolean{false}}
{
\begin{figure}[h!]
\centering
\includegraphics[width=1\textwidth]{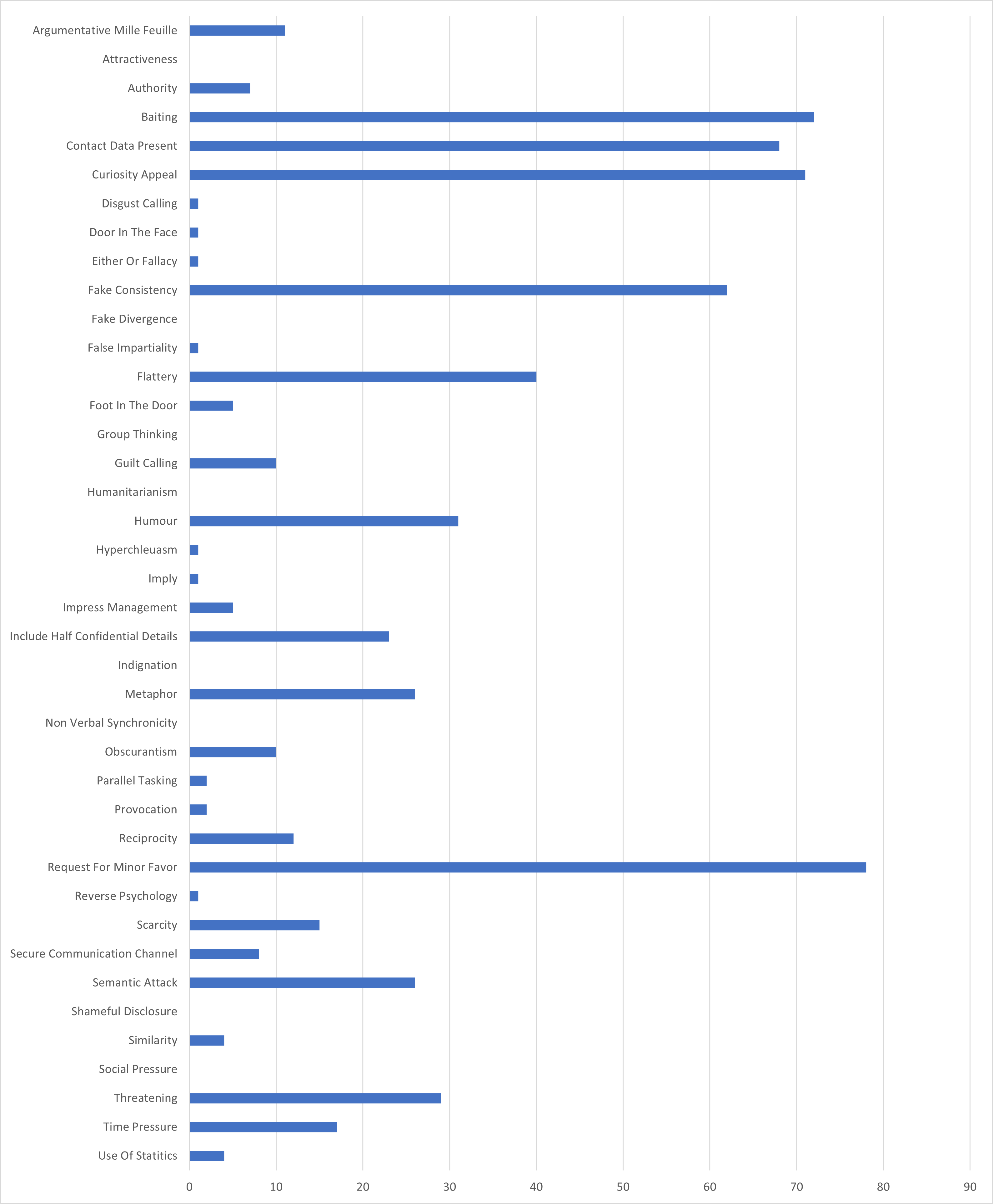}
\caption{Usage in real world of each manipulation technique}
\label{fig:real_world}
\end{figure}
}
As seen in the Table, {\bf Request For Minor Favor} has by far the highest usage with 78\% usage rate, followed by {\bf Baiting} with 72\%, then {\bf Curiosity Appeal} with 71\%. {\bf Contact Data Present} and {\bf Fake Consistency} are the other techniques which have usage over 50\% with respectively 68\% and 62\%. These techniques likely reflect common strategies of piquing interest, offering a tempting "bait," establishing a seemingly legitimate context, or starting with a small, easy request to gain compliance.

One hypothesis is that some of these techniques are inherently less applicable or effective within the typical structure and constraints of phishing emails. For instance, techniques like {\bf Non Verbal Synchronicity} are by definition difficult to convey through text-based communication. Others, such as {\bf Humanitarianism} or {\bf Group Thinking}, might be deemed too overt or less resonant with the individualistic and opportunistic nature of typical phishing attacks. Alternatively, it is possible that these techniques are employed with such subtlety or in contexts not readily apparent in email text alone that they were consistently missed by human annotation. Finally, the zero usage could also reflect a strategic choice by phishers, indicating that these particular manipulation tactics are perceived as less effective in achieving their goals compared to the more prevalent techniques observed in our analysis.

These first results clearly show that some techniques are too few represented both in the training set and in the test set so that we can really conclude about the relevance of our automatic labelization approach for these techniques. As a consequence, in the rest of the paper, for further analyses, we decided to consider only the 21 techniques for which TP+FN > 5.

\section{Detailed Analysis of the 21 selected techniques}

\label{sec:analysissectedtechniques}
\subsection{Performance of the classification}

\ifthenelse{\boolean{false}}
{
\begin{figure}[h!]
\centering
\includegraphics[width=1\textwidth]{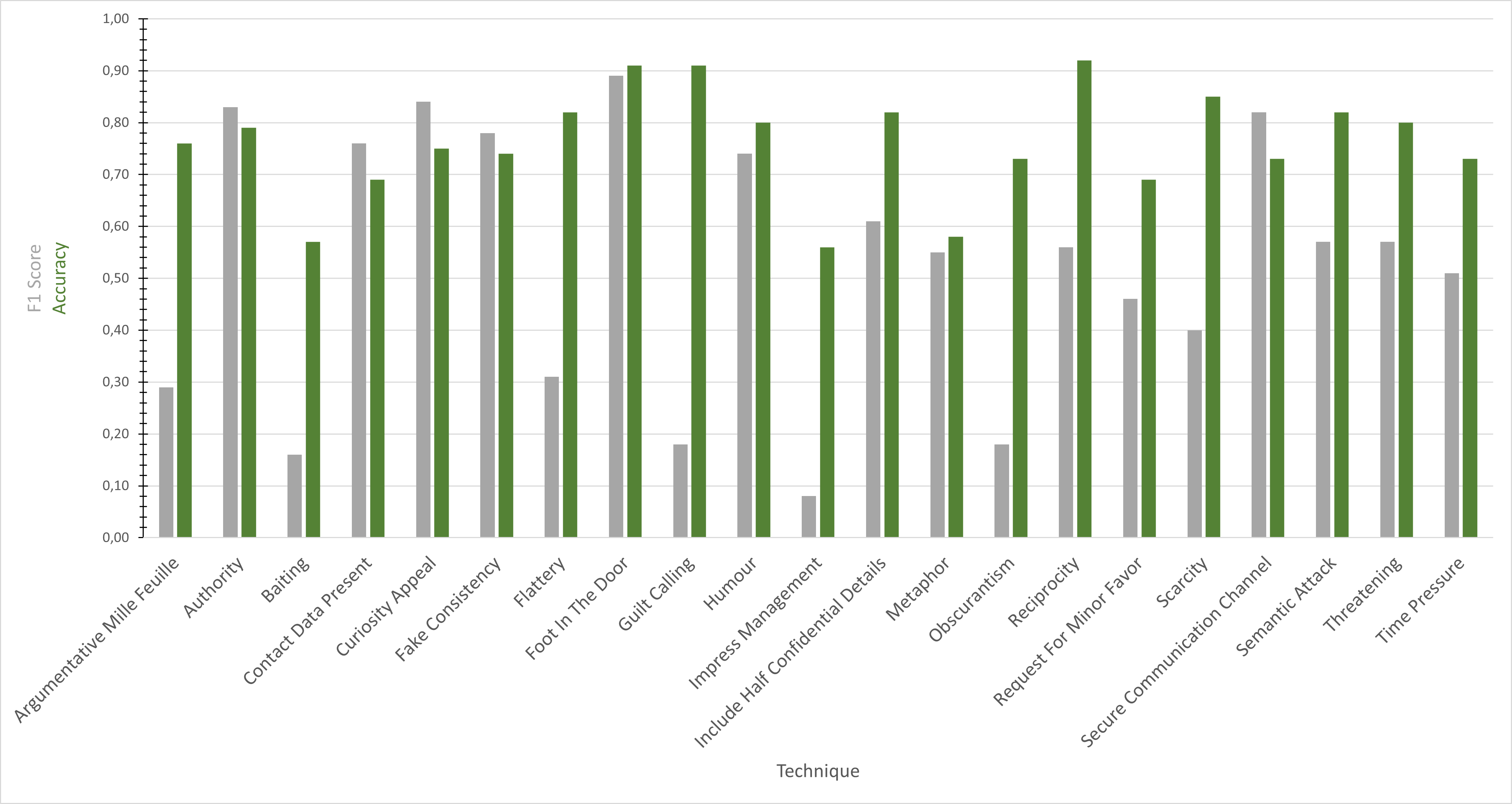}
\caption{F1 Score and accuracy for Each Manipulation Technique}
\label{fig:barchart}
\end{figure}
}

Table \ref{tab:Scoretechniques} provides the F1-score and accuracy
for each of the 21 manipulation techniques and gives a comparative overview of the classification performance across different techniques. The F1-score, representing the harmonic mean of precision and recall, offers a balanced measure of model accuracy. A higher F1-score indicates better overall performance in identifying a specific manipulation technique. Table \ref{tab:Scoretechniques} reveals a wide variation in F1-scores. Several techniques, notably {\bf Request For Minor Favor}, {\bf Baiting}, and {\bf Curiosity Appeal}, exhibit relatively high F1-scores (above 0.8), indicating strong model performance in classifying emails utilizing these tactics.
Techniques like {\bf Argumentative Mille Feuille}, {\bf Authority}, {\bf Foot In The Door}, {\bf Obscurantism}, {\bf Impress Management}, and {\bf Secure Communication Channel} demonstrate lower F1-scores, less than 0.5, representing a significant challenge for accurate classification.
This suggests that the model struggled significantly to correctly classify emails employing these techniques, either failing to identify them when present (low recall) or incorrectly labeling emails as using these techniques when they did not (low precision). One potential contributing factor to these low F1-scores, particularly for techniques with few real examples as shown in Appendix \ref{annex:synthetic}, is the limited representation of these techniques in the original dataset.


The accuracy metric is probably the most relevant metric for our study as it considers all the true labelizations of the LLM (either positive or negative) over the sum of TP+TN+FP+FN. This metric is thus calculated over sufficiently relevant numbers of samples that the F1 (which is sometimes calculated from very little numbers of FP and FN). As our different manipulation techniques are differently represented in the training and test set, the weighted accuracy is an interesting metric as it allows to take into account the imbalance between the different classes of data.     
For each manipulation technique, the accuracy indicates the proportion of times the LLM correctly classifies the emails based on their true nature—whether they include this technique or not. It thus measures how well the LLM detects the presence of a manipulation technique correctly. As can be seen in Table \ref{tab:Scoretechniques}, the accuracy exhibits better results than the F1 score, as the accuracy is always above 0.7, except for {\bf Argumentation Mille Feuille}, {\bf Impress Management} and {\bf Metaphor}. For some manipulation techniques, the LLM exhibit very good results with accuracy superior to 0.8 ({\bf Foot In The Door}, {\bf Flattery}, {\bf Guilt Calling}, {\bf Humour}, {\bf Include Half Confidential Details}, {\bf Reciprocity}, {\bf Secure Communication Channel} and {\bf Threatening} ).
Overall, considering the techniques with "sufficient" representation (TP+FN > 5), the classification approach achieved an Average Weighted Accuracy of 0.76. This metric provides a global view of the performance, weighted by the prevalence of each technique in the test set.  


\subsection{Confusion Matrix}

\begin{figure}[h!]
\centering
\includegraphics[width=0.95\textwidth]{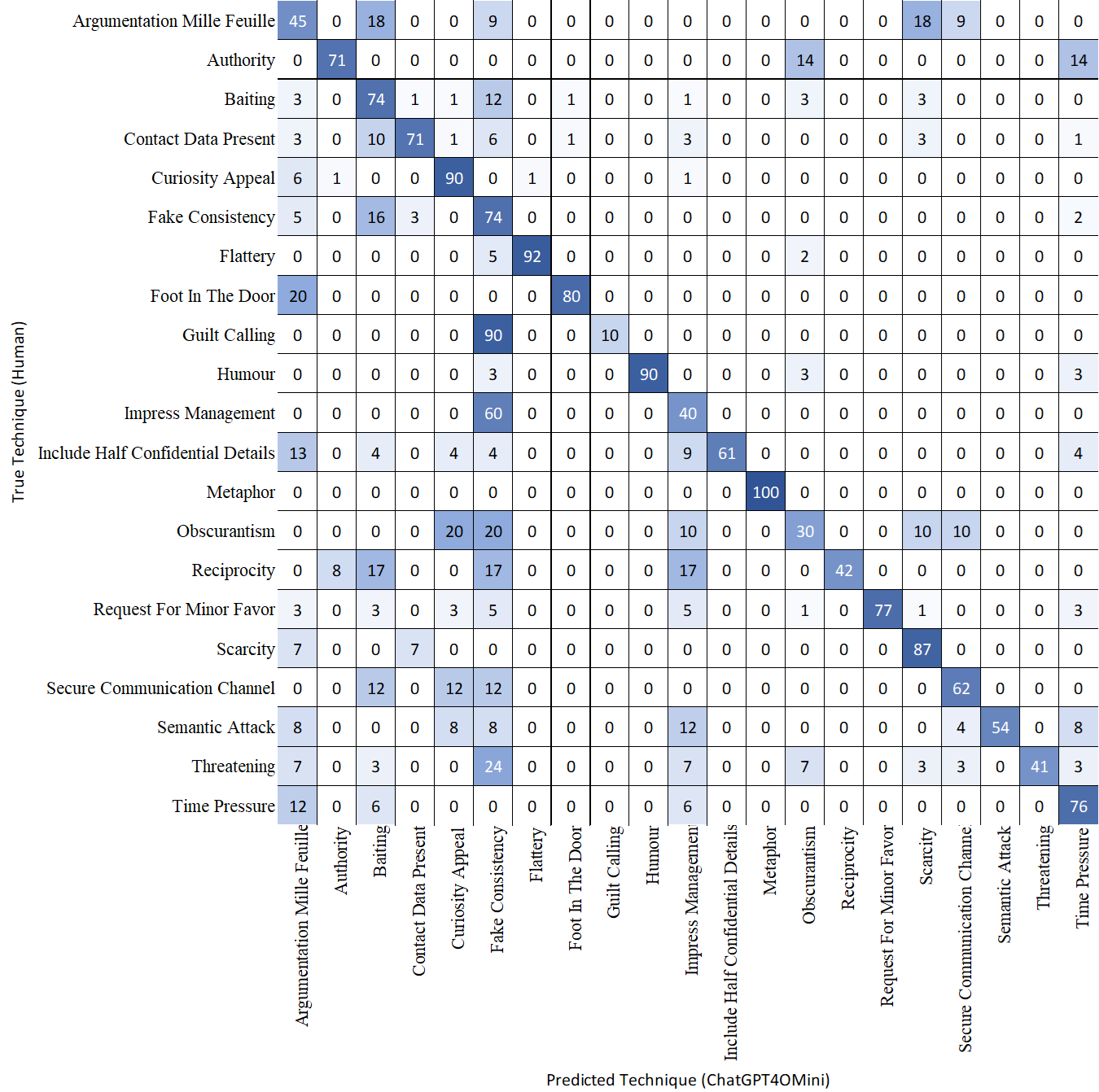}
\caption{Confusion Matrix for the Classification of Phishing Techniques (max score 100)}
\label{fig:confusion_matrix}
\end{figure}

To further understand the performance of this classification method, we identify thanks to the confusion matrix in Figure \ref{fig:confusion_matrix}, the most common confusion errors made by the model. To illustrate how to interpret this matrix, consider the {\bf Metaphor} row. Only a single column contains a non-zero value, 100 indicating perfect classification by the LLM for this technique (each row summing to 100, subject to rounding precision). Conversely, in the {\bf Foot in the Door} row, the LLM misclassifies the technique as {\bf Argumentation Mille-Feuille} approximately 20\% of the time, reflecting a notable confusion between these two categories.

A primary observation from the confusion matrix in Figure \ref{fig:confusion_matrix} is the frequent misclassification of the \textbf{Guilt Calling} technique as \textbf{Fake Consistency} by the model under analysis (GPT-4o-mini), suggesting difficulty in distinguishing between these two categories. This confusion is unlikely to stem from training data artifacts. Firstly, the low co-occurrence observed between these techniques (Figure \ref{fig:co_occurence_matrix}) makes confusion due to frequent joint appearance improbable. Secondly, neither technique involved synthetic data generation (Annex \ref{annex:synthetic}), ruling out issues related to synthetic data representation. A more plausible explanation may lie in the semantic or cognitive similarities between the techniques; both leverage psychological manipulation (inducing guilt versus fostering trust to encourage interaction), and the subtle linguistic cues differentiating them might be insufficiently captured in the LLM's training data. Comparative analysis supports the model-specific nature of this finding: other LLMs tested exhibited different misclassification patterns (e.g., correctly classifying \textbf{Fake Consistency} while misclassifying \textbf{Request For Minor Favor}), confirming that the specific difficulty with \textbf{Guilt Calling} and \textbf{Fake Consistency} is characteristic of GPT-4o-mini. Nevertheless, apart from this specific ambiguity, and particularly when excluding techniques with fewer than five real-world occurrences,
the model demonstrates a commendable overall ability to differentiate between techniques. This robustness extends even to categories with limited data, indicating strong general classification performance despite dataset imbalances.



\ifthenelse{\boolean{false}}
{

\pgfplotstableread[col sep=comma]{
Technique,TP+FN,Accuracy,F1
Authority,9,0.75,0.32
Argumentative Mille Feuille,6,0.48,0.24
Contact Data Present,142,0.58,0.61
Curiosity Appeal,38,0.74,0.83
Fake Consistency,23,0.74,0.76
Foot In The Door,6,0.78,0.27
Flattery,33,0.43,0.58
Guilt Calling,26,0.92,0.43
Humour,41,0.39,0.5
Impress Management,19,0.5,0.07
Include Half Confidential Details,50,0.82,0.61
Metaphor,13,0.58,0.55
Obscurantism,17,0.78,0.21
Reciprocity,0,0.91,0.53
Scarcity,16,0.79,0.55
Secure Communication Channel,7,0.85,0.4
Request For Minor Favor,52,0.77,0.87
Threatening,54,0.83,0.62
Time Pressure,38,0.82,0.62
Baiting,160,0.77,0.83
Semantic Attack,95,0.77,0.53
Disgust Calling,0,0.99,0.67
Door In The Face,0,0.99,0.0
Either Or Fallacy,0,0.99,0.0
False Impartiality,3,0.98,0.5
Hyperchleuasm,0,0.99,0.0
Imply,0,0.99,0.0
Parallel Tasking,3,0.5,0.07
Provocation,1,0.62,0.1
Reverse Psychology,0,0.99,0.0
Similarity,0,0.95,0.62
Use Of Statitics,7,0.89,0.42
Attractiveness,4,0.93,0.0
Group Thinking,0,0.99,0.0
Humanitarianism,0,0.98,0.0
Indignation,1,0.96,0.0
Non Verbal Synchronicity,0,1.0,0.0
Shameful Disclosure,0,0.99,0.0
Social Pressure,2,1.0,0.0
Fake Divergence,0,0.99,0.0
}\datatable

\begin{figure}[htbp]
    \centering
    \begin{tikzpicture}
        \begin{axis}[
            title={F1 Score by Training Set Occurrences},
            xlabel={Training Set Occurrences (Real Examples)},
            ylabel={F1 Score},
            xmin=-5, xmax=165,
            ymin=-0.05, ymax=1.05,
            grid=major,
            legend pos=north west,
            width=0.8\textwidth,
            height=6cm
        ]
        \addplot [
            scatter,
            only marks,
            mark=*,
            mark size=1.5pt,
            red
        ] table [
            x=TP+FN,
            y=F1
        ] {\datatable};
        \addlegendentry{Data Points}

        \addplot [
            no marks,
            thick,
            blue
        ] table [
            y={create col/linear regression={y=F1}},
            x=TP+FN
        ] {\datatable};
        \addlegendentry{Linear Fit}

        \draw [dashed, thick, gray] (axis cs:0, -0.05) -- (axis cs:0, 1.05);
        \draw [dashed, thick, gray] (axis cs:5, -0.05) -- (axis cs:5, 1.05);

        \fill [gray, opacity=0.3] (axis cs:0, -0.05) rectangle (axis cs:5, 1.05);

        \addplot [
            scatter,
            only marks,
            mark=*,
            mark size=1.5pt,
            gray, 
            opacity=0.7 
        ] table [
            x=TP+FN,
            y=F1,
            restrict x to domain=0:5 
        ] {\datatable};

        \node[rotate=90, anchor=south] at (axis cs:2.5, 0.5) {\scriptsize Non representative};

        \end{axis}
    \end{tikzpicture}
    \caption{Classification F1 Score vs. Technique Occurrences in Training Set (Real Examples) with Linear Fit.}
    \label{fig:f1_vs_train_occurrences}
\end{figure}
}


\subsection{Combination of techniques}

\begin{figure}[h!]
\centering
\includegraphics[width=0.95\textwidth]{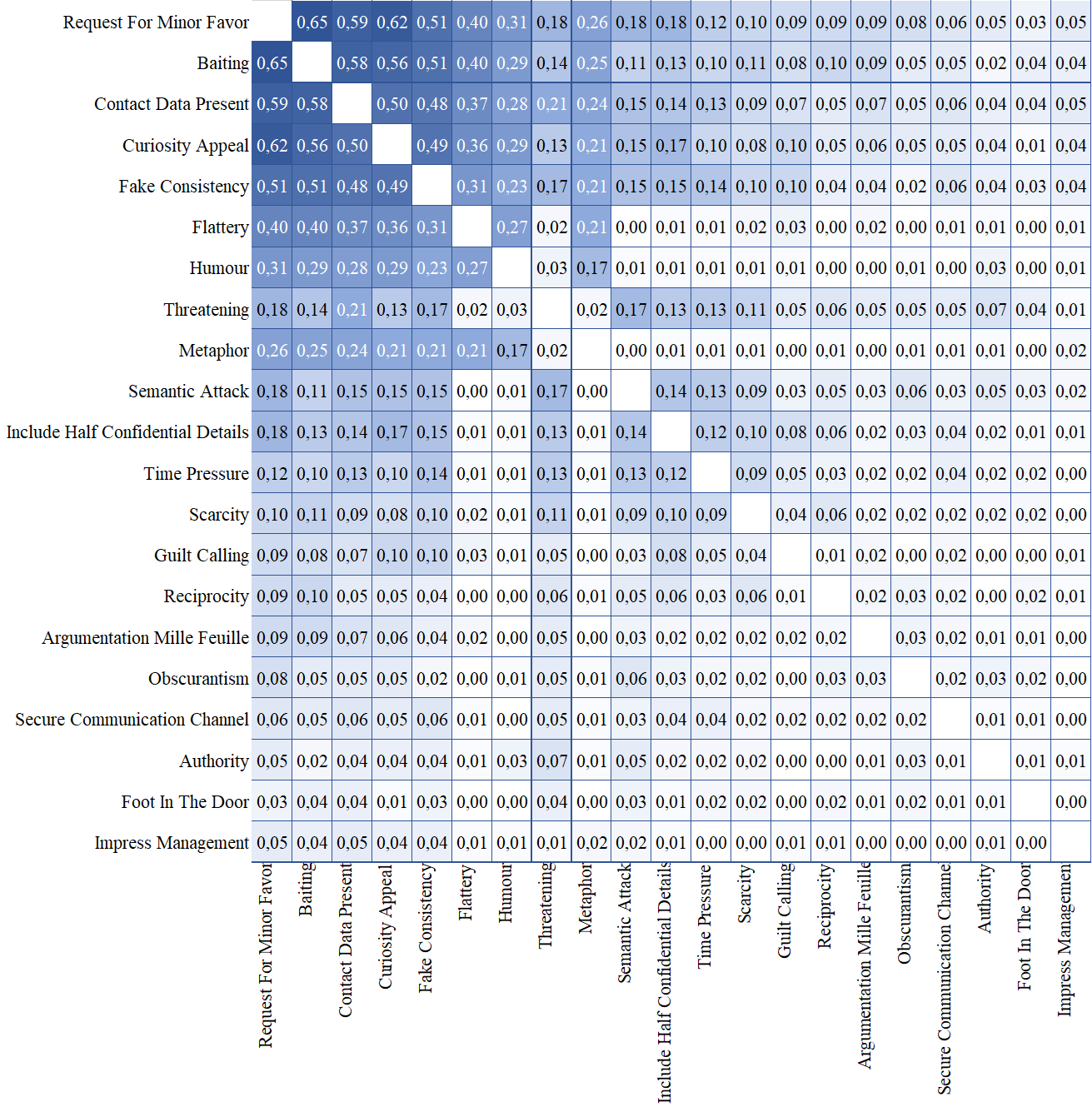}
\caption{Co-occurrence Matrix of Phishing Techniques (normalized)}
\label{fig:co_occurence_matrix}
\end{figure}

Beyond individual technique usage, it's crucial to understand how these techniques are used in combination. Figure \ref{fig:co_occurence_matrix} presents a co-occurrence matrix of the 21 manipulation techniques. Each cell (i, j) in the matrix represents the normalized frequency with which technique i and technique j appear together in the same email. A darker shade of blue indicates a higher co-occurrence frequency. 

Several notable patterns emerge from the co-occurrence matrix. For Example, {\bf Contact Data Present} frequently co-occurs with {\bf Request For Minor Favor} (0.62), {\bf Curiosity Appeal} (0.50) and {\bf Fake Consistency} (0.48). This suggests that providing contact information is often used as a tactic to build trust or legitimacy alongside requests, curiosity-inducing elements, or appeals to consistency. {\bf Baiting} shows strong co-occurrences with {\bf Request For Minor Favor} (0.65) and others. More generally {\bf Contact Data Present}, {\bf Curiosity Appeal}, {\bf Fake Consistency}, {\bf Flattery}, {\bf Request For Minor Favor} and  {\bf Baiting} are the most co-occurring techniques. It means, these techniques are often not used alone.

This usage analysis, combined with the co-occurrence analysis, provides a more accurate and nuanced understanding of which manipulation techniques are most actively employed by phishers, both individually and in combination.

\subsection{Intermodel comparison}
\label{sec:comparison}

To rigorously evaluate the influence of model choice on the efficacy of our In-Context Learning approach, we conducted a comparative analysis across nine leading LLMs. These models were selected on the following criteria: 1) Should not be a reasoning model for fair comparison and 2) One model per leading AI company; open source if available, otherwise closed source; best and most recent model have priority (exception for GPT-4o-mini and GPT-4o).


Performance was measured using the Average Weighted Accuracy metric, computed over the 21 manipulation techniques deemed sufficiently represented in the test set (as discussed in Section \ref{sec:results}). Critically, all models were subjected to identical experimental conditions, employing a consistent ICL prompt structure and a temperature setting of 0. This deterministic configuration (temperature=0) ensures reproducibility and is standard practice for classification tasks where probabilistic sampling is undesirable. 

\begin{figure}[!h]
    \centering
    \includegraphics[width=0.8\textwidth]{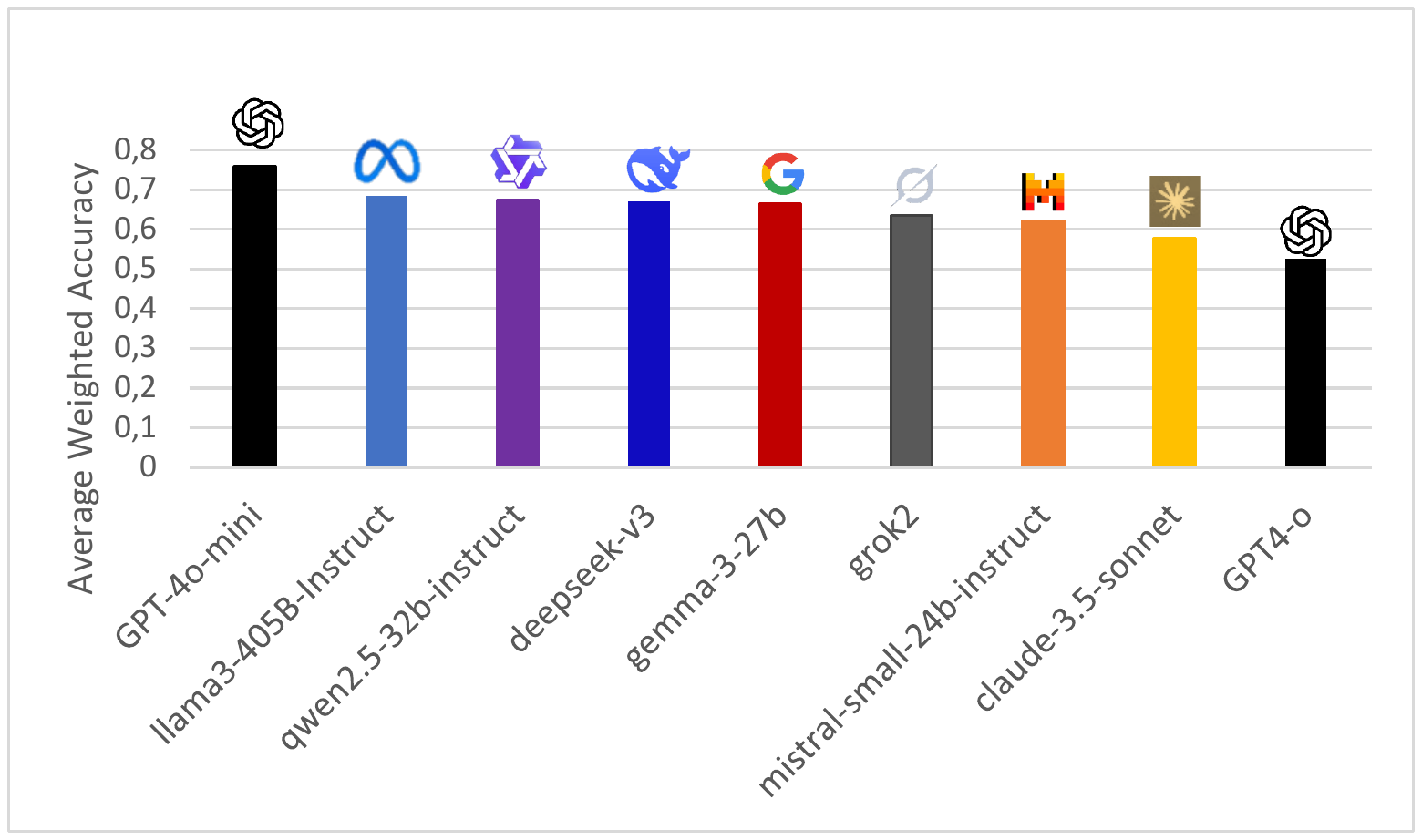}
    \caption{Model performance comparison.}
    \label{fig:intermodel}
\end{figure}

The comparative performance landscape, illustrated in Figure \ref{fig:intermodel}, reveals a notable spread in capabilities for this specific task, with Average Weighted Accuracies ranging from 0.53 to 0.76. GPT-4o-mini distinguished itself as the most effective model, achieving the highest score (0.76), establishing a clear lead over the other evaluated models. Interestingly, several large and capable models formed a tight performance cluster significantly behind the leader: llama3-405B-Instruct (0.68), qwen2.5-32b-instruct (0.67), deepseek-v3 (0.67), and gemma-3-27b (0.66) all scored within the [0.66-0.69] range, indicating robust but secondary performance levels for this classification challenge. Models such as grok2 (0.63) and mistral-small-24b-instruct-2501 (0.62) occupied the middle ground, demonstrating moderate effectiveness. Perhaps surprisingly, the larger GPT4-o model (0.53) significantly underperformed its 'mini' counterpart and ranked last among the tested models, closely followed by claude-3.5-sonnet (0.58). This discrepancy might suggest that for this highly specific, definition-and-example-driven ICL task, the potentially more constrained and instruction-focused nature of GPT-4o-mini proved advantageous over the broader, potentially more complex inference patterns of the full GPT-4o model, or that the larger model is less optimized for this specific type of few-shot classification based on psychological constructs. These findings underscore the critical impact of LLM selection on performance for fine-grained classification tasks using ICL and highlight that model size or general capability do not always directly translate to superior performance in specialized applications like analyzing psychological manipulation techniques in phishing emails.

\section{Discussion}
\label{sec:discussion}

This study's exploration of In-Context Learning (ICL) for phishing email classification, grounded in a granular taxonomy of 21 psychological manipulation techniques, offers several notable strengths. The novelty of applying ICL to this domain, leveraging the capabilities of Large Language Models (LLMs) without explicit fine-tuning, presents a significant methodological advancement. This approach, coupled with the detailed granularity afforded by our taxonomy, provides a more nuanced understanding of persuasive strategies in phishing than prior research. Furthermore, the utilization of a real-world dataset from SignalSpam enhances the relevance and authenticity of our findings, grounding the analysis in contemporary phishing practices. In comparison to traditional technical methodologies focused on surface-level indicators, our content-centric, psychological approach offers a deeper understanding of the underlying manipulation techniques, aligning with the growing recognition of social engineering's importance in phishing attacks. Relative to prior studies examining persuasive elements, our work distinguishes itself through the fine-grained taxonomy and the application of ICL, offering a more adaptable and flexible alternative to traditional machine learning. However, these strengths are juxtaposed with inherent limitations. The dataset size, particularly the human-annotated subset of 100 emails, while valuable for evaluation, inherently restricts the generalizability of performance metrics and the depth of analysis possible for less frequent techniques. As highlighted by the low real-world prevalence of certain techniques and the confusion matrix analysis, a larger annotated corpus would be beneficial for robust conclusions about detectability and error patterns. Moreover, inherent annotation quality concerns, despite our efforts to ensure accuracy through expert review, inevitably introduces a degree of uncertainty. Finally, the computational cost associated with deploying LLMs for ICL presents practical limitations in terms of scalability and widespread implementation. Nevertheless, the findings from this study carry significant practical implications. Our approach paves the way for improved phishing detection tools capable of recognizing a broader spectrum of manipulation techniques, moving beyond purely technical indicators. The detailed taxonomy and prevalence analysis can inform the development of enhanced user awareness training programs, equipping individuals with a more sophisticated understanding of phishers' psychological tactics. Furthermore, the granular identification of manipulation techniques enables the design of targeted interventions to address specific user vulnerabilities and attacker strategies. Finally, the inherent adaptability of ICL offers the potential for creating more responsive and evolving detection systems capable of adapting to the ever-changing landscape of phishing techniques. By bridging the critical gap between technical analysis and the often-overlooked psychological dimensions of phishing, this study contributes towards more effective and comprehensive strategies for combating these pervasive cyber threats.

\section{Conclusion}
\label{sec:conclusion}

This study demonstrated the potential of In-Context Learning with LLMs for the fine-grained classification of psychological manipulation techniques within phishing emails, achieving promising results on a real-world dataset despite limitations in sample size for certain techniques. Our analysis highlighted the prevalence of specific tactics like {\bf Request For Minor Favor}, {\bf Baiting}, and {\bf Curiosity Appeal}, and revealed common co-occurrence patterns, offering valuable insights into attacker methodologies. The comparative analysis of different LLMs underscored the importance of model selection for this specific task, with GPT-4o-mini showing superior performance. Significantly, the demonstrated effectiveness of the ICL classification method unlocks the capability to label phishing data based on psychological techniques at a much larger scale than previously feasible through manual annotation alone. This large-scale labeling, enabled by our approach, in turn unlocks the potential for comprehensive lexical analysis across vast email corpora. Such lexical studies are critical for identifying subtle linguistic markers tied to specific manipulation tactics, thereby paving the way for more sophisticated detection models. Ultimately, this work provides a foundation for developing more psychologically-informed, adaptable, and effective phishing detection and prevention strategies.

\bibliographystyle{splncs04}
\bibliography{full}

\begin{thebibliography}{10}
\providecommand{\url}[1]{\texttt{#1}}
\providecommand{\urlprefix}{URL }
\providecommand{\doi}[1]{https://doi.org/#1}

\bibitem{basnet2011a}
Basnet, R.B., Sung, A.H., Liu, Q.: Rule-based phishing attack detection. In:
  International Conference on Security and Management ({{SAM}} 2011), {{Las
  Vegas}}, {{NV}} (2011)

\bibitem{caputo2014a}
Caputo, D.D., Pfleeger, S.L., Freeman, J.D., Johnson, M.E.: Going {{Spear
  Phishing}}: {{Exploring Embedded Training}} and {{Awareness}}. IEEE Security
  \& Privacy  \textbf{12}(1),  28--38 (Jan 2014). \doi{10.1109/MSP.2013.106}

\bibitem{cialdini2001a}
Cialdini, R.: The science of persuasion. Scientific American  \textbf{284}(2),
  76--81 (2001)

\bibitem{cialdini2009a}
Cialdini, R.B., et~al.: Influence: {{Science}} and Practice, vol.~4. Pearson
  education Boston (2009)

\bibitem{dalmiere:hal-05027416}
Dalmiere, A., Nicomette, V., Auriol, G., Marchand, P.: {A classification of
  manipulation technique used in social engineering attacks and underlying
  cognitive biases, needs, norms, and emotions} (Apr 2025),
  \url{https://hal.science/hal-05027416}, working paper or preprint

\bibitem{dong2020}
Dong, X., Yu, Z., Cao, W., Shi, Y., Ma, Q.: A survey on ensemble learning.
  Frontiers of Computer Science  \textbf{14}(2),  241--258 (Apr 2020).
  \doi{10.1007/s11704-019-8208-z}

\bibitem{evanko2021gpt3}
{Evanko-Douglas}, K.: Can {{GPT-3}} create synthetic training data for machine
  learning models? AmeliorMate Research Report (Jan 2021)

\bibitem{ferreira2015a}
Ferreira, A., Coventry, L., Lenzini, G.: Principles of {{Persuasion}} in
  {{Social Engineering}} and {{Their Use}} in {{Phishing}}. In: Tryfonas, T.,
  Askoxylakis, I. (eds.) Human {{Aspects}} of {{Information Security}},
  {{Privacy}}, and {{Trust}}. pp. 36--47. Lecture {{Notes}} in {{Computer
  Science}}, Springer International Publishing, Cham (2015).
  \doi{10.1007/978-3-319-20376-8_4}

\bibitem{fette2007a}
Fette, I., Sadeh, N., Tomasic, A.: Learning to detect phishing emails. In:
  Proceedings of the 16th International Conference on {{World Wide Web}}. pp.
  649--656. ACM, Banff Alberta Canada (May 2007). \doi{10.1145/1242572.1242660}

\bibitem{fette2007aa}
Fette, I., Sadeh, N., Tomasic, A.: Learning to detect phishing emails. In:
  Proceedings of the 16th International Conference on {{World Wide Web}}. pp.
  649--656. ACM (2007). \doi{10.1145/1242572.1242660},
  \url{https://dl.acm.org/doi/10.1145/1242572.1242660}

\bibitem{garera2007a}
Garera, S., Provos, N., Chew, M., Rubin, A.D.: A framework for detection and
  measurement of phishing attacks. In: Proceedings of the 2007 {{ACM}} Workshop
  on {{Recurring}} Malcode. pp.~1--8. ACM, Alexandria Virginia USA (Nov 2007).
  \doi{10.1145/1314389.1314391}

\bibitem{gikandi2024aa}
Gikandi, J., Kamau, J., Njuguna, D., Sawe, L.: Sentence {{Level Analysis
  Model}} for {{Phishing Detection Using KNN}}  \textbf{6}(1),  25--39 (2024).
  \doi{10.32604/jcs.2023.045859},
  \url{https://www.techscience.com/JCS/v6n1/55181}

\bibitem{goldstein2008a}
Goldstein, N.J., Cialdini, R.B., Griskevicius, V.: A room with a viewpoint:
  {{Using}} social norms to motivate environmental conservation in hotels.
  Journal of consumer Research  \textbf{35}(3),  472--482 (2008)

\bibitem{herley2009a}
Herley, C.: So long, and no thanks for the externalities: The rational
  rejection of security advice by users. In: Proceedings of the 2009 Workshop
  on {{New}} Security Paradigms Workshop. pp. 133--144. ACM, Oxford United
  Kingdom (Sep 2009). \doi{10.1145/1719030.1719050}

\bibitem{jagatic2007a}
Jagatic, T.N., Johnson, N.A., Jakobsson, M., Menczer, F.: Social phishing.
  Communications of the ACM  \textbf{50}(10),  94--100 (Oct 2007).
  \doi{10.1145/1290958.1290968}

\bibitem{kahneman2011a}
Kahneman, D.: Thinking, Fast and Slow. {Farrar, Straus and Giroux}, New York,
  1st ed edn. (2011)

\bibitem{kim2013a}
Kim, D., Hyun~Kim, J.: Understanding persuasive elements in phishing e-mails:
  {{A}} categorical content and semantic network analysis. Online Information
  Review  \textbf{37}(6),  835--850 (Nov 2013). \doi{10.1108/OIR-03-2012-0037}

\bibitem{koide2024aa}
Koide, T., Fukushi, N., Nakano, H., Chiba, D.: {{ChatSpamDetector}}:
  {{Leveraging Large Language Models}} for {{Effective Phishing Email
  Detection}} (2024). \doi{10.48550/ARXIV.2402.18093},
  \url{https://arxiv.org/abs/2402.18093}

\bibitem{kyaw2024aa}
Kyaw, P.H., Gutierrez, J., Ghobakhlou, A.: A {{Systematic Review}} of {{Deep
  Learning Techniques}} for {{Phishing Email Detection}}  \textbf{13}(19),
  ~3823 (2024). \doi{10.3390/electronics13193823},
  \url{https://www.mdpi.com/2079-9292/13/19/3823}

\bibitem{lee2010a}
Lee, K., Caverlee, J., Webb, S.: The social honeypot project: Protecting online
  communities from spammers. In: Proceedings of the 19th International
  Conference on {{World}} Wide Web. pp. 1139--1140. ACM, Raleigh North Carolina
  USA (Apr 2010). \doi{10.1145/1772690.1772843}

\bibitem{li2020a}
Li, Q., Cheng, M., Wang, J., Sun, B.: {{LSTM}} based phishing detection for big
  email data. IEEE Transactions on Big Data  \textbf{8}(1),  278--288 (2020)

\bibitem{long2024}
Long, L., Wang, R., Xiao, R., Zhao, J., Ding, X., Chen, G., Wang, H.: On
  {{LLMs-Driven Synthetic Data Generation}}, {{Curation}}, and {{Evaluation}}:
  {{A Survey}} (Jun 2024). \doi{10.48550/arXiv.2406.15126}

\bibitem{noah2022aa}
Noah, N., Tayachew, A., Ryan, S., Das, S.: Phishercop : Developing an nlp-based
  automated tool for phishing detection  \textbf{66}(1),  2093--2097 (2022).
  \doi{10.1177/1071181322661060},
  \url{https://journals.sagepub.com/doi/10.1177/1071181322661060}

\bibitem{ovelgonne2017a}
Ovelgonne, M., Dumitra{\c s}, T., Prakash, B.A., Subrahmanian, {\relax VS}.,
  Wang, B.: Understanding the relationship between human behavior and
  susceptibility to cyber attacks: {{A}} data-driven approach. ACM Transactions
  on Intelligent Systems and Technology (TIST)  \textbf{8}(4),  1--25 (2017)

\bibitem{pan2024aa}
Pan, F., Wu, X., Li, Z., Luu, A.T.: Are {{LLMs Good Zero-Shot Fallacy
  Classifiers}}? (2024). \doi{10.48550/ARXIV.2410.15050},
  \url{https://arxiv.org/abs/2410.15050}

\bibitem{pratkanis2007a}
Pratkanis, A.R.: Winning hearts and minds: A social influence analysis. In:
  Information Strategy and Warfare, pp. 227--107. Routledge (2007)

\bibitem{salloum2022a}
Salloum, S., Gaber, T., Vadera, S., Shaalan, K.: A {{Systematic Literature
  Review}} on {{Phishing Email Detection Using Natural Language Processing
  Techniques}}  \textbf{10},  65703--65727 (2022).
  \doi{10.1109/ACCESS.2022.3183083},
  \url{https://ieeexplore.ieee.org/document/9795286/}

\bibitem{shahriar2022aa}
Shahriar, S., Mukherjee, A., Gnawali, O.: Improving {{Phishing Detection Via
  Psychological Trait Scoring}} (2022). \doi{10.48550/ARXIV.2208.06792},
  \url{https://arxiv.org/abs/2208.06792}

\bibitem{sheng2010a}
Sheng, S., Holbrook, M., Kumaraguru, P., Cranor, L.F., Downs, J.: Who falls for
  phish?: A demographic analysis of phishing susceptibility and effectiveness
  of interventions. In: Proceedings of the {{SIGCHI Conference}} on {{Human
  Factors}} in {{Computing Systems}}. pp. 373--382. ACM, Atlanta Georgia USA
  (Apr 2010). \doi{10.1145/1753326.1753383}

\bibitem{sheng2007a}
Sheng, S., Magnien, B., Kumaraguru, P., Acquisti, A., Cranor, L.F., Hong, J.,
  Nunge, E.: Anti-{{Phishing Phil}}: The design and evaluation of a game that
  teaches people not to fall for phish. In: Proceedings of the 3rd Symposium on
  {{Usable}} Privacy and Security. pp. 88--99. ACM, Pittsburgh Pennsylvania USA
  (Jul 2007). \doi{10.1145/1280680.1280692}

\bibitem{tian2024a}
Tian, C.A., Jensen, M.L., Bott, G., Luo, X.R.: The influence of affective
  processing on phishing susceptibility. European Journal of Information
  Systems pp. 1--15 (May 2024). \doi{10.1080/0960085X.2024.2351442}

\bibitem{Whitfield_2021}
Whitfield, D.: The magic of synthetic data (Jan 2021)

\bibitem{whittaker2010a}
Whittaker, C., Ryner, B., Nazif, M.: Large-scale automatic classification of
  phishing pages. In: Network and Distributed System Security Symposium (2010)

\bibitem{winata2021}
Winata, G.I., Madotto, A., Lin, Z., Liu, R., Yosinski, J., Fung, P.: Language
  {{Models}} are {{Few-shot Multilingual Learners}}. In: Proceedings of the 1st
  {{Workshop}} on {{Multilingual Representation Learning}}. pp. 1--15.
  Association for Computational Linguistics, Punta Cana, Dominican Republic
  (2021). \doi{10.18653/v1/2021.mrl-1.1}

\bibitem{workman2008a}
Workman, M.: Wisecrackers: {{A}} theory-grounded investigation of phishing and
  pretext social engineering threats to information security. Journal of the
  American Society for Information Science and Technology  \textbf{59}(4),
  662--674 (Feb 2008). \doi{10.1002/asi.20779}

\end{thebibliography}

\appendix  

\section{Psychological manipulation techniques}
\label{annex:techniques}
\ifthenelse{\boolean{false}}
{
\begin{itemize}
    \item \textbf{Attractiveness}: Utilizes physical appeal to build trust rapidly.
    \item \textbf{Authority}: Enhances credibility by associating the message with a seemingly reliable or legitimate source
    \item \textbf{Argumentative Mille Feuille}: Overwhelms the target with a high volume of arguments, regardless of individual argument strength. 
    \item \textbf{Contact Data Present}: Incorporates contact details (phone numbers, email addresses) to falsely signal legitimacy and market orientation.
    \item \textbf{Curiosity Appeal}: Increases  motivation to engage by piquing the target's innate curiosity. 
    \item \textbf{Disgust Calling}: Uses repulsive stimuli (images, descriptions) to induce the emotion of disgust.
    \item \textbf{Door In The Face}: Secures compliance to a smaller, target request by first presenting an unreasonably large, initial request that is likely to be refused.
    \item \textbf{Either Or Fallacy}: Restricts perceived options to a false dichotomy, presenting only two opposing choices.
    \item \textbf{Fake Consistency}: Tricks the target into behaving consistently with a fabricated past or present commitment or belief.
    \item \textbf{Fake Divergence}: Creates an illusion of widespread disagreement or uncertainty on a topic to undermine the target's confidence in their existing knowledge. 
    \item \textbf{False Impartiality}: Presents the manipulator as neutral and disinterested to enhance credibility. 
    \item \textbf{Foot In The Door}: Achieves compliance with a larger, ultimate request by first securing agreement to a smaller, initial request.
    \item \textbf{Group Thinking}: Leverages the desire for group consensus and conformity to suppress dissenting opinions within a group setting. 
    \item \textbf{Flattery}:  Influences by offering compliments and praise. 
    \item \textbf{Guilt Calling}: Manipulates through remorse, exploiting internalized guilt arising from perceived transgressions.
    \item \textbf{Humanitarianism}: Exploits guilt by portraying oneself as a victim in need of help to evoke sympathy and compassion.
    \item \textbf{Hyperchleuasm}: Uses exaggeration or irony to inoculate against reasonable counterarguments by presenting an extreme version.
    \item \textbf{Imply}: Subtly influences through indirect suggestion rather than explicit commands. 
    \item \textbf{Humour}: Distracts attention and reduces message comprehension by employing humor. 
    \item \textbf{Impress Management}: Enhances persuasiveness by projecting positive traits, making the source more appealing and trustworthy during heuristic processing.
    \item \textbf{Indignation}: Evokes moderate anger at perceived injustice or violation of values, increasing motivation to act against the perceived wrongdoer. 
    \item \textbf{Include Half Confidential Details}: Builds a false sense of trust and rapport by including seemingly confidential or personalized details.
    \item \textbf{Metaphor}: Uses figurative language (metaphors) to subtly direct interpretation and frame understanding in a specific, manipulator-preferred way. 
    \item \textbf{Non Verbal Synchronocity}: Enhances message reception and persuasiveness through synchronized non-verbal cues, improving comprehension and agreement during real-time interaction.
    \item \textbf{Obscurantism}: Employs complex jargon and obfuscated language to deceive or impress. 
    \item \textbf{Parallel Tasking}: Reduces attentional capacity and cognitive resources by introducing a secondary task to be performed concurrently with processing the manipulative message.
    \item \textbf{Provocation}: Induces strong anger to heighten motivation to engage, but simultaneously diminishing critical thinking and increasing impulsive behavior directed at the perceived source of provocation. 
    \item \textbf{Reciprocity}:  Obliges the target to reciprocate after receiving a favor or perceived benefit.
    \item \textbf{Reverse Psychology}: Enhances psychological reactance by forbidding an action.
    \item \textbf{Scarcity}: Inflates perceived value and urgency by emphasizing rarity and limited availability. 
    \item \textbf{Secure Communication Channel}: Leverages trusted communication mediums to falsely assure targets of security and legitimacy. 
    \item \textbf{Shameful Disclosure}: Manipulates using the threat of revealing embarrassing or norm-violating behavior. 
    \item \textbf{Similarity}: Establishes perceived shared traits or characteristics to build trust and rapport rapidly.
    \item \textbf{Social Pressure}: Influences by demonstrating that others (can be a fabricated majority) are already complying or adopting a certain behavior. 
    \item \textbf{Request For Minor Favor}: Seeks small, seemingly harmless acts of compliance initially (micro-commitments).
    \item \textbf{Threatening}: Instills fear by conveying potential negative consequences. 
    \item \textbf{Use Of Statistics}: Employs seemingly objective numerical data and statistics to sway opinions and create a sense of certainty and validity.
    \item \textbf{Time Pressure}: Reduces cognitive processing and critical evaluation time by creating a sense of urgency and time sensitivity. 
    \item \textbf{Baiting}: Uses positive emotional stimuli, such as offers of rewards, money, or romantic promises, to lure targets and direct their attention.
    \item \textbf{Semantic Attack}: Employs visually similar domain names, URLs, or email addresses to mimic legitimate entities and deceive users. 
\end{itemize}
}
\ifthenelse{\boolean{false}}
{
{\scriptsize
    \begin{longtable}{|m{3cm}|p{9cm}|} 
        \hline
        \centering  Argumentative Mille Feuille & Overwhelms the target with a high volume of arguments, regardless of individual argument strength. \\
        \hline
        \centering Attractiveness & Utilizes physical appeal to build trust rapidly. \\
        \hline
        \centering Authority & Enhances credibility by associating the message with a seemingly reliable or legitimate source. \\
        \hline
        \centering Baiting & Uses rewards or emotional stimuli to lure targets. \\
        \hline
        \centering Contact Data Present & Incorporates contact details (phone numbers, email addresses) to falsely signal legitimacy and market orientation. \\
        \hline
        \centering Curiosity Appeal & Increases motivation to engage by piquing the target's innate curiosity. \\
        \hline
        \centering  Disgust Calling & Uses repulsive stimuli (images, descriptions) to induce the emotion of disgust. \\
        \hline
        \centering  Door In The Face & Secures compliance to a smaller, target request by first presenting an unreasonably large, initial request that is likely to be refused. \\
        \hline
        \centering Either Or Fallacy & Restricts perceived options to a false dichotomy, presenting only two opposing choices. \\
        \hline
        \centering Fake Consistency & Tricks the target into behaving consistently with a fabricated past or present commitment or belief. \\
        \hline
        \centering Fake Divergence & Creates an illusion of widespread disagreement or uncertainty on a topic to undermine the target's confidence. \\
        \hline
        \centering  False Impartiality & Presents the manipulator as neutral and disinterested to enhance credibility. \\
        \hline
        \centering Foot In The Door & Achieves compliance with a larger, ultimate request by first securing agreement to a smaller, initial request. \\
        \hline
        \centering Flattery & Influences by offering compliments and praise. \\
        \hline
        \centering Group Thinking & Leverages the desire for group consensus and conformity to suppress dissenting opinions. \\
        \hline
        \centering Guilt Calling & Manipulates through remorse, exploiting internalized guilt arising from perceived transgressions. \\
        \hline
        \centering Humanitarianism & Exploits guilt by portraying oneself as a victim in need of help. \\
        \hline
        \centering Hyperchleuasm & Uses exaggeration or irony to inoculate against reasonable counterarguments. \\
        \hline
        \centering Humour & Distracts attention and reduces message comprehension by employing humor. \\
        \hline
        \centering Imply & Subtly influences through indirect suggestion rather than explicit commands. \\
        \hline
        \centering Impress Management & Enhances persuasiveness by projecting positive traits. \\
        \hline
        \centering Include Half Confidential Details & Builds a false sense of trust by including seemingly confidential or personalized details. \\
        \hline
        \centering Indignation & Evokes moderate anger at perceived injustice, increasing motivation to act. \\
        \hline
        \centering  Metaphor & Uses figurative language to direct interpretation and frame understanding. \\
        \hline
        \centering  Non Verbal Synchronocity & Enhances persuasiveness through synchronized non-verbal cues. \\
        \hline
        \centering  Obscurantism & Employs complex jargon and obfuscated language to deceive or impress. \\
        \hline
        \centering  Parallel Tasking & Reduces cognitive resources by introducing a secondary task. \\
        \hline
        \centering Provocation & Induces anger to increase impulsive behavior and diminish critical thinking. \\
        \hline
        \centering Reciprocity & Obliges the target to reciprocate after receiving a favor. \\
        \hline
        \centering Request For Minor Favor & Seeks micro-commitments through small, harmless requests. \\
        \hline
        \centering Reverse Psychology & Enhances reactance by forbidding an action. \\
        \hline
        \centering Scarcity & Inflates perceived value by emphasizing rarity and limited availability. \\
        \hline
        \centering Secure Communication Channel & Leverages trusted channels to falsely assure security. \\
        \hline
        \centering Semantic Attack & Mimics legitimate entities using similar names or URLs. \\
        \hline
        \centering Shameful Disclosure & Manipulates using the threat of revealing embarrassing behavior. \\
        \hline
        \centering Similarity & Establishes perceived shared traits to build trust. \\
        \hline
        \centering Social Pressure & Demonstrates that others are complying to influence behavior. \\
        \hline
        \centering Threatening & Instills fear by conveying potential negative consequences. \\
        \hline
        \centering Time Pressure & Reduces critical evaluation by creating urgency. \\
        \hline
        \centering Use Of Statistics & Uses numerical data to create a sense of certainty. \\
        \hline
    \end{longtable}
}
}

{\scriptsize
    \begin{longtable}{|m{3cm}|p{9cm}|} 
        \hline
        \centering  Argumentative Mille Feuille & Overwhelms the target with a high volume of arguments, regardless of individual argument strength. \\
        \hline
        \centering Authority & Enhances credibility by associating the message with a seemingly reliable or legitimate source. \\
        \hline
        \centering Baiting & Uses rewards or emotional stimuli to lure targets. \\
        \hline
        \centering Contact Data Present & Incorporates contact details (phone numbers, email addresses) to falsely signal legitimacy and market orientation. \\
        \hline
        \centering Curiosity Appeal & Increases motivation to engage by piquing the target's innate curiosity. \\
        \hline
        \centering Fake Consistency & Tricks the target into behaving consistently with a fabricated past or present commitment or belief. \\
        \hline
        \centering Foot In The Door & Achieves compliance with a larger, ultimate request by first securing agreement to a smaller, initial request. \\
        \hline
        \centering Flattery & Influences by offering compliments and praise. \\
        \hline
        \centering Guilt Calling & Manipulates through remorse, exploiting internalized guilt arising from perceived transgressions. \\
        \hline
        \centering Humour & Distracts attention and reduces message comprehension by employing humor. \\
        \hline
        \centering Impress Management & Enhances persuasiveness by projecting positive traits. \\
        \hline
        \centering Include Half Confidential Details & Builds a false sense of trust by including seemingly confidential or personalized details. \\
        \hline
        \centering  Metaphor & Uses figurative language to direct interpretation and frame understanding. \\
        \hline
        \centering  Obscurantism & Employs complex jargon and obfuscated language to deceive or impress. \\
        \hline
        \centering Reciprocity & Obliges the target to reciprocate after receiving a favor. \\
        \hline
        \centering Request For Minor Favor & Seeks micro-commitments through small, harmless requests. \\
        \hline
        \centering Scarcity & Inflates perceived value by emphasizing rarity and limited availability. \\
        \hline
        \centering Secure Communication Channel & Leverages trusted channels to falsely assure security. \\
        \hline
        \centering Semantic Attack & Mimics legitimate entities using similar names or URLs. \\
        \hline
        \centering Threatening & Instills fear by conveying potential negative consequences. \\
        \hline
        \centering Time Pressure & Reduces critical evaluation by creating urgency. \\
        \hline
    \end{longtable}
}

\section{Number of synthetic examples for manipulation techniques lacking real-life examples}
\label{annex:synthetic}
\begin{multicols}{2}
\begin{itemize}
    \item \textbf{Attractiveness}: 1
    \item \textbf{Disgust Calling}: 5
    \item \textbf{Door in the Face}: 5
    \item \textbf{Either or Fallacy}: 5
    \item \textbf{Fake Divergence}: 5
    \item \textbf{False Impartiality}: 2
    \item \textbf{Group Thinking}: 5
    \item \textbf{Humanitarianism}: 5
    \item \textbf{Hyperchleuasm}: 5
    \item \textbf{Imply}: 5
    \item \textbf{Indignation}: 4
    \item \textbf{Non-Verbal Synchronicity}: 5
    \item \textbf{Parallel Tasking}: 2
    \item \textbf{Provocation}: 4
    \item \textbf{Reciprocity}: 5
    \item \textbf{Reverse Psychology}: 5
    \item \textbf{Shameful Disclosure}: 5
    \item \textbf{Similarity}: 5
    \item \textbf{Social Pressure}: 3
\end{itemize}
\end{multicols}

\ifthenelse{\boolean{false}}
{
\begin{itemize}
    \item \textbf{Attractiveness}: 1
    \item \textbf{Disgust Calling}: 5
    \item \textbf{Door in the Face}: 5
    \item \textbf{Either or Fallacy}: 5
    \item \textbf{Fake Divergence}: 5
    \item \textbf{False Impartiality}: 2
    \item \textbf{Group Thinking}: 5
    \item \textbf{Humanitarianism}: 5
    \item \textbf{Hyperchleuasm}: 5
    \item \textbf{Imply}: 5
    \item \textbf{Indignation}: 4
    \item \textbf{Non-Verbal Synchronicity}: 5
    \item \textbf{Parallel Tasking}: 2
    \item \textbf{Provocation}: 4
    \item \textbf{Reciprocity}: 5
    \item \textbf{Reverse Psychology}: 5
    \item \textbf{Shameful Disclosure}: 5
    \item \textbf{Similarity}: 5
    \item \textbf{Social Pressure}: 3
\end{itemize}
}
\ifthenelse{\boolean{false}}
{
\begin{table}[!ht]
\centering
\caption{List of techniques with their number of real-life examples and synthetic mails}
\label{tab:techniques} 
\begin{tabular}{lcc}
\toprule
\textbf{Technique name} & \textbf{Real examples} & \textbf{Synthetic mails} \\
\midrule
Argumentative Mille Feuille & 6 & 0 \\
Attractiveness & 4 & 1 \\
Authority & 9 & 0 \\
Baiting & 160 & 0 \\
Contact Data Present & 142 & 0 \\
Curiosity Appeal & 38 & 0 \\
Disgust Calling & 0 & 5 \\
Door in the Face & 0 & 5 \\
Either or Fallacy & 0 & 5 \\
Fake Consistency & 23 & 0 \\
Fake Divergence & 0 & 5 \\
False Impartiality & 3 & 2 \\
Flattery & 33 & 0 \\
Foot in the Door & 6 & 0 \\
Group Thinking & 0 & 5 \\
Guilt Calling & 26 & 0 \\
Humanitarianism & 0 & 5 \\
Humour & 41 & 0 \\
Hyperchleuasm & 0 & 5 \\
Imply & 0 & 5 \\
Impress Management & 19 & 0 \\
Include Half Confidential Details & 50 & 0 \\
Indignation & 1 & 4 \\
Metaphor & 13 & 0 \\
Non-Verbal Synchronicity & 0 & 5 \\
Obscurantism & 17 & 0 \\
Parallel Tasking & 3 & 2 \\
Provocation & 1 & 4 \\
Reciprocity & 0 & 5 \\
Request for Minor Favor & 52 & 0 \\
Reverse Psychology & 0 & 5 \\
Scarcity & 16 & 0 \\
Secure Communication Channel & 7 & 0 \\
Semantic Attack & 95 & 0 \\
Shameful Disclosure & 0 & 5 \\
Similarity & 0 & 5 \\
Social Pressure & 2 & 3 \\
Threatening & 54 & 0 \\
Time Pressure & 38 & 0 \\
Use of Statistics & 7 & 0 \\
\midrule
\textbf{Total} & \textbf{866} & \textbf{81} \\
\bottomrule
\end{tabular}
\end{table}
}

\ifthenelse{\boolean{false}}
{
\begin{table}[!ht]
\centering
\caption{List of techniques involved in synthetic mails}
\label{tab:techniques} 
\begin{tabular}{lcc}
\toprule
\textbf{Technique name} & \textbf{Synthetic mails} \\
\midrule
Attractiveness  & 1 \\
Disgust Calling  & 5 \\
Door in the Face  & 5 \\
Either or Fallacy  & 5 \\
Fake Divergence  & 5 \\
False Impartiality  & 2 \\
Group Thinking  & 5 \\
Humanitarianism  & 5 \\
Hyperchleuasm  & 5 \\
Imply  & 5 \\
Indignation  & 4 \\
Non-Verbal Synchronicity & 5 \\
Parallel Tasking  & 2 \\
Provocation  & 4 \\
Reciprocity  & 5 \\
Reverse Psychology  & 5 \\
Shameful Disclosure  & 5 \\
Similarity  & 5 \\
Social Pressure  & 3 \\
\midrule
\end{tabular}
\end{table}
}

\section{Prompt for synthetic data generation}
\label{annex:prompt}

Generate 5 examples of mail using {{technique name}}. Don't include template or generic element in mail but create fictions name. Each example must be separate by '\#\#\#'. Mail object is prefixed by Mail Object: and body is prefixed MLLM\_TEXT:. Write all in plaintext block, not markdown (not bold or anything). Every 3 examples maximum (not each example, only if relevant), include in MLLM\_TEXT an image by enclosing their short visual description between brackets.
Here are the definition of this technique :
{{Technique academic definition}}

\section{System prompt}
\label{annex:systempromptclassification}
\begin{framed}
\noindent
\textbf{User prompt:}
\noindent
You are a classifier that responds only with \texttt{'YES'} or \texttt{'NO'}. You help cybersecurity researchers to classify manipulation techniques in emails.

\noindent
\texttt{\{\{Academical definition of the technique\}\}}

\noindent
\textbf{Example :}

\noindent
\texttt{\{\{Example 1\}\}}

\noindent
\#\#\#

\noindent
\texttt{\{\{Example 2\}\}}

\noindent
\#\#\#

\noindent
\texttt{\{\{Example 3\}\}}

\noindent
\#\#\#

\noindent
\texttt{\{\{Example 4\}\}}

\noindent
\#\#\#

\noindent
\textbf{Content to analyze:}

\noindent
\textbf{Mail Object:} \texttt{\{\{Mail Object\}\}}

\noindent
\textbf{MLLM\_TEXT:} \texttt{\{\{Mail content\}\}}

\noindent
\textbf{Attachments:} \texttt{\{\{Attachment filenames\}\}}
\end{framed}

\end{document}